\documentclass[10pt,amsmath,amssymb,nofootinbib,twoside,twocolumn,superscriptaddress,floats,floatfix,aps,prd,preprintnumbers]{revtex4-2}

\usepackage{graphicx,epsfig}
\usepackage{amssymb,amsmath,amsthm,amsfonts}
\usepackage{eufrak}
\usepackage{mathrsfs}
\usepackage{bm}
\usepackage[inline]{enumitem}
\usepackage{tensor}
\usepackage[linktocpage]{hyperref}
\usepackage[capitalise]{cleveref}
\usepackage[usenames,dvipsnames]{xcolor}
\usepackage{url}
\usepackage[inline]{enumitem}
\usepackage{xspace}
\usepackage{comment}
\usepackage{ dsfont }
\usepackage[normalem]{ulem}
\usepackage{subfigure}
\hypersetup{
    colorlinks=true,
    linkcolor=red,
    citecolor=blue,
    filecolor=black,
    urlcolor=blue,
    pdfauthor={F. Corelli, P. Pani, and A. P. Sanna},
    pdftitle={}
}

\newcommand{\gint}{\int_\Omega \text{d}^4x \, \sqrt{-g} \,}    
\newcommand{\tder}{\partial_{t}}                        
\newcommand{\rder}{\partial_{r}}                        

\newcommand{\dd}{\text{d}}                              

\newcommand{\rH}{r_\text{H}}                            
\newcommand{\rS}{r_\text{S}}                            
\newcommand{\rE}{r_\text{E}}                            

\newcommand{\M}{\widetilde{M}}                          

\newcommand{\MBH}{M_\text{BH}}                          

\newcommand{\RR}{\mathcal{R}}                           
\newcommand{\GG}{\mathcal{G}}                           
\newcommand{\OO}[1]{\mathcal{O}\left(#1\right)}         

\newcommand{\cpp}{C\nolinebreak\hspace{-.05em}\raisebox{.4ex}{\tiny\bf +}\nolinebreak\hspace{-.10em}\raisebox{.4ex}{\tiny\bf +}}

\newcommand{\sapienza}{Dipartimento di Fisica, Sapienza Università 
	di Roma, Piazzale Aldo Moro 5, 00185, Roma, Italy}
\newcommand{\infn}{INFN, Sezione di Roma, Piazzale Aldo Moro 2, 00185, Roma, Italy}

\allowdisplaybreaks

\begin{document}

\author{Fabrizio~Corelli}
\email{fabrizio.corelli@uniroma1.it}
\affiliation{\sapienza}\affiliation{\infn}

\author{Paolo~Pani}
\email{paolo.pani@uniroma1.it}
\affiliation{\sapienza}\affiliation{\infn}

\author{Andrea~P.~Sanna}
\email{asanna@roma1.infn.it}
\affiliation{\infn}

\title{Gravity with higher-curvature terms and second-order field equations: \\$f(\RR)$ meets Gauss-Bonnet
}

\begin{abstract}
General Relativity is expected to break down in the high-curvature regime. Beyond an effective field theory treatment with higher-order operators, it is important to identify consistent theories with higher-curvature terms at the nonperturbative level. Two well-studied examples are $f(\RR)$ gravity and Einstein-dilaton-Gauss-Bonnet (EdGB) gravity. The former shares the same vacuum solutions as General Relativity, including black holes, while the latter suffers from well-posedness issues due to quadratic curvature terms in the strong-coupling regime. 
We show that combining these two theories leads to genuinely new phenomena beyond their simple superposition. The resulting framework falls outside Horndeski’s class, as it can be recast as a gravitational theory involving two nonminimally coupled scalar fields with nontrivial mutual interactions. This construction naturally extends EdGB gravity to include arbitrary higher-curvature terms, providing a versatile setting to address fundamental questions. 
Focusing on quadratic and quartic corrections, we find that: (i) black holes are modified by $f(\RR)$ terms, unlike the case without Gauss–Bonnet interactions;  (ii) the resulting solutions retain the qualitative nonperturbative features of EdGB black holes with certain couplings, such as a minimum mass and multiple branches; (iii) a nontrivial mechanism suppresses the divergence of the Ricci scalar in the black-hole interior; (iv) still, even with quartic corrections, the singularity structure and elliptic regions inside the horizon remain similar to those of pure EdGB gravity. This suggests that, at the nonperturbative level, the theory’s ill-posedness cannot be resolved by adding individual higher-order terms. 
This conjecture could be tested by studying the nonlinear dynamics, which remains governed by second-order field equations. 
\end{abstract}

\maketitle

\section{Introduction}
\label{sec:Intro}

Despite its formidable success across a wide range of scales, it is widely believed that General Relativity~(GR) must eventually break down in the high-curvature regime. Its ultraviolet~(UV) completion should resolve the curvature singularities that inevitably form as a generic outcome of gravitational collapse within GR~\cite{Penrose:1964wq}, while also improving its quantum behavior.

A famous attempt to improve the UV behavior of gravity was made in~\cite{Stelle:1976gc} by adding quadratic curvature terms to the Einstein-Hilbert action. Remarkably, the resulting theory is renormalizable, but at the cost of introducing ghost instabilities. The latter arise from the extra degrees of freedom associated with higher-order derivatives in the field equations. This provides further motivation to search for theories that remain healthy while modifying the UV regime of gravity.

A necessary condition for a healthy theory is the absence of Ostrogradsky’s instability, which requires that the field equations be of second differential order~\cite{Ostrogradsky}. Within the context of gravity coupled to a \emph{single} scalar field, the most general class of scalar–tensor theories with second-order field equations is Horndeski's class~\cite{Horndeski:1974wa,Deffayet:2011gz}. This framework is therefore the natural starting point to identify nonpathological modifications of the UV regime of gravity. The only two representatives of such theories are:
\begin{itemize}
    \item $f(\RR)$ theories, where the action contains a generic function of the Ricci scalar $\RR$ (see, e.g.,~\cite{Sotiriou:2008rp} for a review). These can be recast as certain Brans–Dicke theories and hence are included in Horndeski's classification;
    \item Einstein-dilaton-Gauss-Bonnet~(EdGB) theory, wherein the action contains an additional scalar field coupled to the Gauss–Bonnet quadratic curvature invariant~\cite{Gross:1986mw}. The special combination of derivatives entering the Gauss–Bonnet term ensures that the field equations remain second order, while the scalar coupling prevents the Gauss-Bonnet term from being a topological invariant and hence with no impact on the field equations.
\end{itemize}
When considered \emph{individually}, $f(\RR)$ gravity and EdGB theory are the only members of Horndeski's class featuring higher-curvature terms. The main goal of this paper is to explore a gravitational theory containing \emph{both} contributions, which we refer to as $f(\RR)$-dGB gravity (see action~\eqref{eq:Action} below).

Interestingly, when these terms are combined, the resulting theory no longer belongs to Horndeski's class. Indeed, as we shall show, $f(\RR)$–dGB gravity can be reformulated as a theory of gravity with \emph{two} nonminimally coupled scalar fields featuring mutual interactions\footnote{This theory differs from that considered in~\cite{Liu:2020yqa}; there the approach is to start with $f(\RR)=\RR+\alpha \RR^2$, recast it as a Brans–Dicke scalar–tensor theory, and couple the \emph{same} scalar field to the Gauss–Bonnet term. The resulting theory in that case still belongs to Horndeski's class.} and, as such, it belongs to a bi-scalar extension of Horndeski's theory~\cite{Horndeski:2024hee} (see also the model in~\cite{Eichhorn:2023iab} for another example of such class).  It is therefore qualitatively different from a simple superposition of $f(\RR)$ and EdGB theories.

Our focus will be on the UV behavior of this theory, for which black-hole~(BH) solutions provide the natural testing ground.
Although $f(\RR)$ gravity can include arbitrarily high powers of the Ricci curvature in the action, it shares all Ricci-constant solutions of GR, including the vacuum ones.
Moreover, under reasonable assumptions on the form of the $f(\RR)$ function, BHs in $f(\RR)$ gravity coincide with those in GR~\cite{Sotiriou:2011dz}. Consequently, $f(\RR)$ theories are particularly relevant in cosmological~\cite{Starobinsky:1980te} or matter-dominated settings, but remain trivial as regards the resolution of BH singularities.

In contrast, EdGB gravity has been extensively studied for its impact on BH solutions~\cite{Kanti:1995vq,Mignemi:1992nt} and, more recently, as a testbed to investigate BH dynamics beyond GR in a fully nonlinear regime~\cite{Ripley:2019aqj,Ripley:2019irj,Kovacs:2020ywu,East:2020hgw,Witek:2020uzz,East:2021bqk,AresteSalo:2023mmd,Corman:2022xqg,Corelli:2022pio,Corelli:2022phw,Corman:2024cdr}.
However, despite having second-order field equations, the resulting dynamical system is not always hyperbolic: elliptic spacetime regions can form during the evolution of regular initial data~\cite{Ripley:2019aqj,Ripley:2019hxt,Ripley:2019irj,Ripley:2020vpk,East:2021bqk,Corelli:2022pio,Corelli:2022phw,R:2022hlf,Doneva:2023oww,AresteSalo:2025sxc}.
This has motivated the study of theories supplementing EdGB gravity with additional terms (such as~\cite{Antoniou:2021zoy}), which could possibly restore hyperbolicity~\cite{Liu:2022fxy,Thaalba:2023fmq,Thaalba:2024htc,Thaalba:2024crk,Doneva:2024ntw}. Although such modifications can mitigate the problem in some cases, none of the proposed couplings has been shown to eliminate it entirely. This might be due to the fact that the coupling considered in~\cite{Liu:2022fxy,Thaalba:2023fmq,Thaalba:2024htc,Thaalba:2024crk,Doneva:2024ntw} is linear in the Ricci curvature, and therefore subleading in the high-curvature regime when compared to the Gauss-Bonnet term. A different approach to cure the loss of hyperbolicity by adding higher-order differential operators has been developed in~\cite{Figueras:2024bba,Figueras:2025gal}, but is rooted in an effective field theory expansion of the beyond-GR terms. 

In this context, $f(\RR)$-dGB gravity is particularly promising, as it modifies EdGB gravity through terms of arbitrarily high order in the Ricci curvature, which are expected to dominate in the UV regime, without the need of an effective-field-theory expansion.
Furthermore, when added to EdGB theory, the $f(\RR)$ term also affects vacuum solutions such as BHs, since the latter are not Ricci-constant in EdGB gravity. Thus, $f(\RR)$–dGB gravity provides an ideal framework to investigate UV deviations from GR, assess their potential role in resolving BH singularities, and examine the well-posedness of the corresponding dynamical system.

Despite the potential for significant changes in the high-curvature regime, we shall show that a nontrivial  mechanism forces the Ricci scalar (which becomes a dynamical degree of freedom in this theory) to remain finite near the BH singularity. As a result, $f(\RR)$ corrections in the BH interior are always mild, and one qualitatively recovers the same singularity and elliptic region found inside BHs in EdGB gravity.

The rest of the paper is structured as follows. In \cref{sec:Theory&FE}, we present the theory and the field equations. We then focus on static, asymptotically-flat, BH solutions in \cref{subsec:StaticBHSolutions}, describing the boundary conditions and the numerical scheme used to evaluate the solutions (\cref{subsubsec:BoundaryConditions} and \cref{subsubsec:NumericalIntegration}, respectively). \Cref{sec:Results} contains the main results of the numerical integration. We describe the general behavior of the BH solutions (\cref{subsec:SolutionsBehavior}), the domain of existence (\cref{subsec:ExistenceDomain}), their internal structure (\cref{subsec:internal}) and, finally, the hyperbolicity and well-posedness issues (\cref{subsec:HyperbolicityWellPosedness}). We finally draw our conclusions in \cref{sec:Conclusions}.

As the explicit expressions of the field equations, for both the static and time-dependent cases, are quite lengthy, we do not reproduce them here. They are, instead, provided in an ancillary \textsc{Mathematica} notebook.

Throughout the paper, we adopt units in which $c = G = 1$.

\section{Theory and field equations}
\label{sec:Theory&FE}

We consider an action of the form
\begin{equation}
    S = \frac{1}{16\pi}\gint \biggl\{f(\RR) - \bigl(\nabla_\mu \phi\bigr)\bigl(\nabla^\mu \phi\bigr)  + 2 \eta(\phi) \GG\biggr\}\, ,
    \label{eq:Action}
\end{equation}
where $f(\RR)$ is a generic function of the Ricci scalar $\RR$, and $\phi$ is a scalar field coupled to the Gauss-Bonnet invariant $\GG = \RR^2 - 4 \mathcal{R}^{\mu\nu}\mathcal{R}_{\mu\nu} + \mathcal{R}^{\mu\nu\rho\sigma}\mathcal{R}_{\mu\nu\rho\sigma}$ through the coupling function $\eta(\phi)$. 
The theory represents a combination of $f(\RR)$ gravity and EdGB theory. The 
latter can be framed as a low-energy truncation of string theory, in which case the scalar is a dilaton field and the coupling $\eta(\phi)$ is exponential~\cite{Gross:1986mw}.
We refer to the theory in Eq.~\eqref{eq:Action} as $f(\RR)$-dGB gravity. As we shall discuss, this theory does not represent a mild modification of EdGB, but rather an entirely different theory.

When combined as in Eq.~\eqref{eq:Action}, these terms give rise to a theory that lies outside Horndeski’s class~\cite{Horndeski:1974wa,Deffayet:2011gz}. In fact, $f(\RR)$–dGB gravity can be recast as a scalar–tensor theory involving two nonminimally coupled scalar fields, by exploiting the well-known equivalence between $f(\RR)$ gravity and certain scalar-tensor theories~\cite{Sotiriou:2008rp}. One approach to recast the former into the latter was derived in Ref.~\cite{Jaime:2010kn}, and consists in identifying the Ricci curvature\footnote{In the more standard approach, one identifies the functional derivative of $f$ as the scalar field~\cite{Sotiriou:2008rp}.} with an additional scalar field $\chi$. Following their procedure, the action~\eqref{eq:Action} can be re-written as (throughout the paper, unless otherwise specified, the prime will denote derivation with respect to the function's argument)
\begin{align}
    S = \frac{1}{16\pi}\gint \biggl\{& f'(\chi) \RR - V(\chi) - \bigl(\nabla_\mu \phi\bigr)\bigl(\nabla^\mu \phi\bigr) \notag\\
    &+ 2 \eta(\phi) \GG\biggr\}\, ,
    \label{eq:Action_JPS}
\end{align}
where the scalar potential is defined as
\begin{equation}
    V(\chi) \equiv f'(\chi) \,  \chi - f(\chi)\, .
    \label{eq:VChiGeneral}
\end{equation}
The field equations derived from \cref{eq:Action_JPS} are
\begin{align}
    f''(\chi) \bigl( \RR - \chi) &= 0,  \label{eq:FieldPsiJPS} \\
    \Box \phi &= -\eta'(\phi) \GG, \label{eq:FieldDilatonJPS} \\
    G_{\mu\nu} &= \frac{8 \pi}{f'(\chi)} T_{\mu\nu}^\text{dGB} - \frac{1}{2 f'(\chi)} V(\chi) g_{\mu\nu} \notag\\
    &+ \frac{f'''(\chi)}{f'(\chi)} \Bigl[ \bigl( \nabla_\mu \chi \bigr) \bigl( \nabla_\nu \chi \bigr) - g_{\mu\nu} \bigl( \nabla \chi \bigr)^2 \Bigr] \notag \\
    &+ \frac{f''(\chi)}{f'(\chi)} \Bigl[ \nabla_\mu \nabla_\nu \chi - g_{\mu\nu} \Box \chi \Bigr], \label{eq:FieldGravJPS}
\end{align}
where $\bigl( \nabla \chi )^2 = \bigl(\nabla_\lambda \chi \bigr) \bigl(\nabla^\lambda \chi \bigr)$, $\Box = \nabla_\mu \nabla^\mu$, while $T_{\mu\nu}^\text{dGB}$ is the stress-energy tensor associated to the EdGB sector in \cref{eq:Action_JPS}:
\begin{align}
    T_{\mu\nu}^\text{dGB} =& \frac{1}{8\pi} \biggl[ \bigl( \nabla_\mu \phi \bigr) \bigl( \nabla_\nu \phi) - \frac{1}{2} \bigl(\nabla_\alpha \phi\bigr)\bigl(\nabla^\alpha \phi\bigr) g_{\mu\nu} \notag \\ 
           &- 2\bigl( \nabla_\gamma\nabla^\alpha \eta(\phi) \bigr) \delta^{\gamma\delta\kappa\lambda}_{\alpha\beta\rho\sigma} \tensor{\RR}{^{\rho\sigma}_{\kappa\lambda}} \tensor{\delta}{^\beta_{(\mu}} \tensor{g}{_{\nu) \delta}}\biggr]\, ,
   \label{eq:TmunusGB}
\end{align}
with $\delta^{\gamma\delta\kappa\lambda}_{\alpha\beta\rho\sigma}$ being the generalized Kronecker delta. 
Whenever $f''(\chi)\neq0$, Eq.~\eqref{eq:FieldPsiJPS} enforces $\RR=\chi$, so the Ricci curvature becomes a dynamical degree of freedom, as in the standard $f(\RR)$ gravity~\cite{Sotiriou:2008rp,Jaime:2010kn}. The field equation for $\RR=\chi$ can be obtained by taking the trace of Eq.~\eqref{eq:TmunusGB}, namely
\begin{equation}
   \Box\chi =\frac{1}{3 f''} \left[8\pi T^{\rm dGB}-3(\nabla\chi)^2\, f'''   +2 f-\chi\, f'\right]\,,\label{trace}
\end{equation}
where we also used Eq.~\eqref{eq:VChiGeneral}. If $T^{\rm dGB}=0$, from the above equation we see that $\chi=\RR={\rm const.}$ is a solution provided that $2 f=\chi\, f'$ --- which is an algebraic equation for $\RR$ for a given $f(\RR)$ --- admits at least a solution. Indeed, in the absence of matter all Ricci-constant GR solutions are also solutions to $f(\RR)$ gravity~\cite{Sotiriou:2008rp}. This includes the family of Kerr-Newman BHs with possibly a cosmological constant. However, this is not the case here, since generically $T^{\rm dGB}\neq0$. Hence, BH solutions in $f(\RR)$-dGB are bound to have a nontrivial profile of the extra $\chi$ field and are generally different from their EdGB counterpart. 

A specific $f(\RR)$-dGB theory is selected once the dilatonic coupling $\eta(\phi)$ and the form of $f(\RR)$ are specified. In this paper, we consider an exponential dilatonic coupling~\cite{Gross:1986mw}
\begin{equation}
    \eta(\phi) = \lambda e^{-\gamma \phi},
    \label{eq:DilatonicCoupling}
\end{equation}
where $\gamma$ is the dimensionless dilaton coupling constant, while $\lambda$ is the Gauss-Bonnet coupling constant with dimensions of a squared length. With such an exponential coupling, the theory described by the action~\eqref{eq:Action_JPS} shares with the EdGB theory the invariance under the simultaneous transformations 
\begin{equation}
    \phi \to \phi + C\, , \quad \lambda \to \lambda e^{\gamma C}\, ,
    \label{eq:symmetrylambda}
\end{equation}
with $C$ a real constant. 

Concerning the specific form of the $f(\RR)$, in the following we consider the family
\begin{equation}
    f(\RR) = \RR + \kappa \RR^n\, , \quad n \in \mathds{N}
    \label{eq:fRGeneral}
\end{equation}
where $\kappa = \ell^{2(n-1)}$ is a coupling constant and $\ell$ the corresponding reference length scale. For concreteness, in \cref{sec:Results}, we consider the cases with $n = 2$ and $n = 4$, which will be also referred to as ``quadratic" and ``quartic" case, respectively. The quadratic case includes a term that is in principle as important as the Gauss-Bonnet term in the high-curvature regime, while the quartic case should in principle be dominant in the same regime. Therefore, within $f(\RR)$-dGB gravity we can drastically modify the dynamics of EdGB theory while retaining second-order field equations.

\subsection{Static BH solutions}
\label{subsec:StaticBHSolutions}

In this paper, we construct static, asymptotically-flat, spherically-symmetric BH solutions of the theory~\eqref{eq:Action_JPS}. A key focus of our analysis will be to examine the hyperbolicity of the system of equations, including the interior regions of these objects, where EdGB alone is known to be problematic~\cite{Ripley:2019aqj,Ripley:2019irj,East:2021bqk,Corelli:2022phw,Corelli:2022pio,R:2022hlf}. For this purpose, we primarily employ horizon-penetrating, i.e., Painlevé-Gullstrand (PG)-like, coordinates $(t, \, r, \, \theta, \, \varphi)$. In the static case, the line element reads as
\begin{equation}
    \dd s^2 = -\alpha(r)^2 \, \dd t^2 + \left[\dd r + \alpha(r) \, \zeta(r) \dd t \right]^2 + r^2 \dd \Omega_2^2\, ,
    \label{eq:LineElementPG}
\end{equation}
while the two scalar fields, $\phi$ and $\chi$, are functions of the radial coordinate $r$ only. Plugging the ansatz~\eqref{eq:LineElementPG} into the field equations~\eqref{eq:FieldPsiJPS}-\eqref{eq:FieldGravJPS} and after some manipulation, we obtain a system of differential equations that has the following form
\begin{align}
    \frac{\alpha'(r)}{\alpha(r)} & = \frac{\mathscr{N}_\alpha \bigl(r, \, \zeta, \, \phi, \, \phi', \, \chi, \, \chi' \bigr)}{\mathscr{D}_\alpha \bigl(r, \, \zeta, \, \phi, \, \phi', \, \chi, \, \chi' \bigr)} \, ; \label{eq:Diffalpha}\\
    \zeta'(r) & = \frac{\mathscr{N}_\zeta \bigl(r, \, \zeta, \, \phi, \, \phi', \, \chi, \, \chi' \bigr)}{\mathscr{D}_\zeta \bigl(r, \, \zeta, \, \phi, \, \phi', \, \chi, \, \chi' \bigr)} \, ; \label{eq:DiffZeta}\\
    \phi''(r) & = \frac{\mathscr{N}_\phi \bigl(r, \, \zeta, \, \phi, \, \phi', \, \chi, \, \chi' \bigr)}{\bigl(\zeta^2 - 1 \bigr) \, \mathscr{D}_\phi \bigl(r, \, \zeta, \, \phi, \, \phi', \, \chi, \, \chi' \bigr)} \, ; \label{eq:DiffPhi}\\
    \chi''(r) & = \frac{\mathscr{N}_\chi \bigl(r, \, \zeta, \, \phi, \, \phi', \, \chi, \, \chi' \bigr)}{\bigl(\zeta^2 - 1 \bigr) \, \mathscr{D}_\chi \bigl(r, \, \zeta, \, \phi, \, \phi', \, \chi, \, \chi' \bigr)}\, . \label{eq:DiffChi}
\end{align}
We first note that $\alpha$ is defined only up to a multiplicative constant, reflecting the freedom to rescale the time coordinate. Moreover, \cref{eq:DiffZeta,eq:DiffPhi,eq:DiffChi} are independent of $\alpha$ and can therefore be integrated to obtain $\zeta$, $\phi$, and $\chi$. The remaining equation, \cref{eq:Diffalpha}, can then be solved separately.

As a final remark, we notice that, for  $\kappa \to 0$, in which \cref{eq:fRGeneral} reduces to the standard Einstein-Hilbert term, \cref{eq:FieldPsiJPS} becomes an identity. However, as we shall see, this equation plays a crucial role in determining the $f(\RR)$-dGB solutions. As a result, the EdGB BH solutions cannot be obtained merely by solving Eqs.~\eqref{eq:Diffalpha}–\eqref{eq:DiffChi} with $\kappa = 0$, and the $\kappa\to0$ limit must be taken a posteriori.

\subsubsection{Boundary conditions and regularity at the horizon}
\label{subsubsec:BoundaryConditions}

To uniquely determine BH solutions, we supplement the system~\eqref{eq:Diffalpha}–\eqref{eq:DiffChi} with boundary conditions imposed both at spatial infinity ($r \to \infty$) and at the horizon $\rH$.

At $r \to \infty$, we require the solutions to be asymptotically flat. 
Following Ref.~\cite{Ripley:2019aqj}, we define the total mass $\MBH$ as the limiting value of the Misner–Sharp mass:
\begin{equation}
    \MBH \equiv \lim_{r\to \infty} M_\text{MS}(r) = \lim_{r\to \infty} \frac{r}{2} \zeta(r)^2\, .
    \label{eq:MBH}
\end{equation}
Expanding the metric functions and the scalar fields in powers of $1/r$, the asymptotic solution reads as (these leading terms are the same for $n = 2, 4$ in~\eqref{eq:fRGeneral}) 
\begin{align}
    \alpha(r) &= A \left[1-\frac{D^2}{4r^2} - \frac{2 D^2 \MBH}{3 r^3} \right] + \mathcal{O}\left(\frac{1}{r^4} \right) \, , \label{eq:alphaEpxansionInf} \\
    \phi(r) &= \phi_\infty + \frac{D}{r} + \frac{D \MBH}{r^2} \notag \\
            &- \frac{D(D^2 - 16 \MBH^2)}{12 r^3} + \mathcal{O}\left(\frac{1}{r^4} \right)\, , \label{eq:phiExpansionInf}\\
    \zeta(r) &= \sqrt{\frac{2\MBH}{r}}-\frac{D^2}{4\sqrt{2\MBH}}\frac{1}{r^{3/2}} \notag \\
             &- \frac{D^2(D^2 + 16 \MBH^2)}{64 \MBH \sqrt{2\MBH}}\frac{1}{r^{5/2}} + \mathcal{O}\left(\frac{1}{r^{7/2}} \right)\, . \label{eq:zetaExpansionInf}
\end{align}
Here, $D$ denotes the scalar ``charge'', while $\phi_\infty$ and $A$ are arbitrary constants. The former can be set to zero by exploiting the theory's symmetry under the transformations~\eqref{eq:symmetrylambda}, whereas $A$ can be normalized to unity by an appropriate rescaling of the time coordinate. The asymptotic behavior of $\chi$ depends on the choice of the functional form of $f(\RR)$. Restricting to the family in Eq.~\eqref{eq:fRGeneral}, we get
\begin{equation}
    \chi(r) = \frac{D^2}{r^4} + \frac{2 D^2 \MBH}{r^5} + \frac{4 D^2 (\MBH^2 + 18 \kappa)}{r^6} + \OO{\frac{1}{r^7}}
    \label{eq:chiExpansionInfQuadratic}
\end{equation}
for the case $n = 2$, and 
\begin{equation}
    \chi(r) = \frac{D^2}{r^4} + \frac{2 D^2 \MBH}{r^5} + \frac{4 D^2 \MBH^2}{r^6} + \OO{\frac{1}{r^7}}
    \label{eq:chiExpansionInfQuartic}
\end{equation}
for the case $n = 4$.
\\

At the horizon we impose regularity of the metric functions and scalar fields. Since \cref{eq:Diffalpha,eq:DiffZeta} are first-order equations, only $\alpha(\rH)$ and $\zeta(\rH)$ are required as boundary conditions. The horizon is identified by $\zeta(\rH)=1$. Since $\alpha$ is defined only up to a multiplicative constant (see \cref{eq:Diffalpha}), its horizon value can be fixed by requiring consistency with the asymptotic behavior~\eqref{eq:alphaEpxansionInf}.

The remaining equations, \cref{eq:DiffPhi,eq:DiffChi}, are of second differential order; therefore, both the values and the derivatives of the scalar fields at the horizon must be specified. Regularity at $\rH$ then imposes a behavior of the form
\begin{align}
    \phi(r) &\simeq \phi_{\text{H}0} + \phi_{\text{H}1}(r-\rH) + \mathcal{O}(r-\rH)^2\, ,\\
    \chi(r) & \simeq \chi_{\text{H}0} + \chi_{\text{H}1}(r-\rH) + \mathcal{O}(r-\rH)^2\, ,
\end{align}
where $\phi_{\text{H}0} \equiv \phi(\rH)$, $\phi_{\text{H}1} \equiv \partial_r\phi(r)|_{r = \rH}$, $\chi_{\text{H}0} \equiv \chi(\rH)$, $\chi_{\text{H}1} \equiv \partial_r\chi(r)|_{r = \rH}$. 
Inspection of \cref{eq:DiffPhi,eq:DiffChi} shows that their right-hand sides diverge at the horizon due to the factor $(\zeta^2 - 1)^{-1}$. Therefore, imposing regularity requires that both
\begin{align}
    \mathscr{N}_\phi\bigl(\rH, \, \zeta = 1, \, \phi_{\text{H}0}, \phi_{\text{H}1}, \, \chi_{\text{H}0}, \, \chi_{\text{H}1}\bigr) &= 0\, ,\label{eq:NphiReg}\\
    \mathscr{N}_\chi\bigl(\rH, \, \zeta = 1, \, \phi_{\text{H}0}, \phi_{\text{H}1}, \, \chi_{\text{H}0}, \, \chi_{\text{H}1}\bigr) &= 0 \label{eq:NchiReg}
\end{align}
are satisfied simultaneously. Solving the system directly is, however, very complicated as it is of third degree in $\phi_{\text{H}1}$ and $\chi_{\text{H}1}$, and the two equations are coupled. Nevertheless, if we start from the equation $\chi = \RR$, by removing the derivatives of the metric in the Ricci scalar via the field equations, and evaluating it at the horizon, we obtain a quadratic equation for $\phi_{\text{H}1}$, which can be solved algebraically. This yields
\begin{equation}
    \phi^{(\pm)}_{\text{H}1} = \frac{\mathfrak{a}_1 \pm \mathfrak{a}_2 \sqrt{\Delta_\text{H}}}{\mathfrak{a}_3}\, ,
    \label{eq:phiH1roots}
\end{equation}
where we defined
\begin{align}
    \mathfrak{a}_1 & \equiv -64 \rH^2 \chi_{\text{H}0} \chi_{\text{H}1} f''(\chi _{\text{H}0}) \eta'(\phi_{\text{H}0})^2 \notag \\
    &+128 \chi_{\text{H}1} f''(\chi_{\text{H}0}) \eta' (\phi_{\text{H}0})^2\notag\\
    &+\rH^6 \chi_{\text{H}1} V(\chi_{\text{H}0})  f''(\chi_{\text{H}0})\notag \\
    &-128 \rH \chi_{\text{H}0} f'(\chi_{\text{H}0})  \eta'(\phi_{\text{H}0})^2\notag \\
    &+2 \rH^5 V(\chi_{\text{H}0})  f'(\chi_{\text{H}0})-4 \rH^3  f'(\chi_{\text{H}0})^2\notag \\
    &-2 \rH^4 \chi_{\text{H}1}  f'(\chi_{\text{H}0})  f''(\chi_{\text{H}0})\notag \\
    &+128 \rH V(\chi_{\text{H}0})  \eta'(\phi_{\text{H}0})^2\, ;\\
    \mathfrak{a}_2 & \equiv \frac{2  f'(\chi_{\text{H}0})-\rH^2 V(\chi_{\text{H}0})}{\rH}\, ;\\
    \mathfrak{a}_3 & \equiv 8 \eta' \left(\phi _{\text{H}0}\right) \biggl[4 \rH^2 f'(\chi _{\text{H}0})\notag\\
    &+\frac{64 \left(\rH^2 \chi_{\text{H}0}-2\right)  \eta'(\phi_{\text{H}0})^2}{\rH^2}-2 \rH^4 V(\chi_{\text{H}0})\biggr]\, ;\\
    \Delta_\text{H} & \equiv -256 \rH^5 \chi_{\text{H}1}  f''(\chi_{\text{H}0})  \eta'(\phi_{\text{H}0})^2\notag\\
    &+\rH^{10} \, \chi_{\text{H}1}^2  f''(\chi_{\text{H}0})^2+4 \rH^8  f'(\chi_{\text{H}0})^2\notag\\
    &+4 f'(\chi_{\text{H}0}) \left[\rH^9 \chi_{\text{H}1}  f''(\chi_{\text{H}0})-192 \rH^4 \eta' (\phi_{\text{H}0})^2\right]\notag\\
    &-4096 \rH^2 \chi_{\text{H}0}  \eta'(\phi_{\text{H}0})^4+128 \rH^6 V(\chi_{\text{H}0})  \eta'(\phi_{\text{H}0})^2 \notag\\
    &+24576 \eta'(\phi_{\text{H}0})^4\, .
\end{align}
Our numerical investigations indicate that asymptotically-flat BH solutions arise in the $\phi^{(+)}_{\text{H}1}$ branch, while we have not found such solutions for $\phi^{(-)}_{\text{H}1}$. The existence of solutions associated with the latter branch remains an interesting possibility, similarly to what happens in EdGB theory (see, e.g., Refs.~\cite{Kanti:1995vq,Kleihaus:2019rbg,Kleihaus:2020qwo}). 

We then substitute the expression of $\phi^{(+)}_{\text{H}1}$ into \cref{eq:NphiReg,eq:NchiReg}, that now become equations in $\chi_{\text{H}0}$, $\chi_{\text{H}1}$ and $\phi_{\text{H}0}$:
\begin{align}
    \mathscr{N}_\phi\bigl(\rH, \, \zeta = 1, \, \phi_{\text{H}0}, \, \chi_{\text{H}0}, \, \chi_{\text{H}1}\bigr) &= 0\, , \label{eq:NphiNophi1}\\
    \mathscr{N}_\chi\bigl(\rH, \, \zeta = 1, \, \phi_{\text{H}0}, \, \chi_{\text{H}0}, \, \chi_{\text{H}1}\bigr) &= 0  \, .\label{eq:NchiNophi1}
\end{align}
Solving them directly is still cumbersome, but can be done numerically, yielding a value for $\chi_{\text{H}1}$ that depends on $\chi_{\text{H}0}$ and $\phi_{\text{H}0}$. Replacing it into Eq.~\eqref{eq:phiH1roots}, we can finally obtain $\phi_{\text{H}1}$. In summary, regularity at the horizon imposes conditions that fix the values of $\chi_{\text{H}1}$ and $\phi_{\text{H}1}$, so that the only independent parameters are $\chi_{\text{H}0}$ and $\phi_{\text{H}0}$.

Exploiting the shift symmetry~\eqref{eq:symmetrylambda}, the value of $\phi_{\text{H}0}$ can initially be chosen arbitrarily. After evaluating the corresponding solution, one can reapply \cref{eq:symmetrylambda}, fixing the constant $C$ so that the $\phi$ solution matches~\eqref{eq:phiExpansionInf} at spatial infinity. The final value of $\phi_{\text{H}0}$ is then determined accordingly.

\subsubsection{Numerical integration}
\label{subsubsec:NumericalIntegration}

Constructing a BH metric consists in finding a solution to Eqs.~\eqref{eq:DiffZeta}-\eqref{eq:DiffChi} that satisfies both the boundary conditions at the horizon and the asymptotic behaviors discussed in the previous section, for some suitable choices of $\chi_{\text{H}0}$ and $\phi_{\text{H}0}$ at the horizon or, equivalently, for the corresponding asymptotic parameters $\MBH$, $D$, and $\phi_\infty$. We perform this task using a shooting method that proceeds in the following way.

First, defining a length scale $\M$, we set the position of the horizon radius at $\rH = 2 \M$ (in practice, we adopt units such that $\rH = 2$), and we set the values of the constants $\lambda$ and $\kappa$, which appear in Eqs.~\eqref{eq:DilatonicCoupling} and~\eqref{eq:fRGeneral}, respectively. Then, we fix the values of $\chi_{\text{H}0}$ and $\phi_{\text{H}0}$, solve numerically Eq.~\eqref{eq:NchiNophi1} for $\chi_{\text{H}1}$, and determine the derivative of $\phi$ at the horizon using $\phi^{(+)}_{\text{H}1}$ in Eq.~\eqref{eq:phiH1roots}. In order to ensure that  Eq.~\eqref{eq:NphiNophi1} is also satisfied at the horizon, we performed the same operations starting from it and verified that the proposed solutions for $\chi_{\text{H}1}$ agree within a relative accuracy of one part in $10^{30}$. 

With the values of $\chi_{\text{H}0}$, $\phi_{\text{H}0}$, $\chi_{\text{H}1}$ and $\phi_{\text{H}1}$, as well as $\zeta_{\text{H}} = 1$, we can now proceed with the numerical integration of Eqs.~\eqref{eq:DiffZeta}-\eqref{eq:DiffChi}. However, while the boundary conditions at the horizon guarantee regularity of the right-hand-sides of Eqs.~\eqref{eq:DiffPhi} and~\eqref{eq:DiffChi}, their denominators will still vanish, and therefore they cannot be evaluated exactly at $r=\rH$ in the integration algorithm. To overcome this issue, we start the integration from an inner boundary $r = \rH + \epsilon$, using
\begin{align}
    \phi(\rH + \epsilon) &= \phi_{\text{H}0} + \epsilon ~ \phi_{\text{H}1} \, , \notag \\
    \chi(\rH + \epsilon) &= \chi_{\text{H}0} + \epsilon ~ \chi_{\text{H}1} \, , \notag \\
    \zeta(\rH + \epsilon) &= 1 +  \epsilon ~ \zeta'(\rH) \, , \notag \\
    \phi'(\rH + \epsilon) &= \phi_{\text{H}1} \, , \notag \\
    \chi'(\rH + \epsilon) &= \chi_{\text{H}1} \, 
    \label{eq:BCHorizonNumerical}
\end{align}
as initial data, where $\zeta'(\rH)$ is obtained from Eq.~\eqref{eq:DiffZeta}, and $\epsilon$ is set to $10^{-10} \, \M$. Note that in principle expanding $\phi'$ and $\chi'$ at order zero can spoil the regularity at the horizon by introducing errors; nevertheless, this does not seem to happen, and our numerical solutions appear continuous. Furthermore, we verified the numerical stability of the results under a change of $\epsilon$.

We terminate the numerical integration at a given outer boundary $r = r_\infty$. Here we wish to evaluate the discrepancy between our solution and the expected asymptotic behaviors; the values of $\chi_{\text{H}0}$ and $\phi_{\text{H}0}$ are fixed by minimizing this discrepancy. For this purpose, we estimate the BH mass and scalar charge as
\begin{equation}
    \MBH = \frac{r_\infty}{2} \zeta(r_\infty)^2, \quad D = - r_\infty^2 \rder \phi \rvert_{r = r_\infty} ,
    \label{eq:MDAsymptoticEstimate}
\end{equation}
which allow us to compute the violation of the asymptotic behavior for $\chi$. As for the dilaton, we still need to handle the parameter $\phi_\infty$, which also affects the final value of $\phi_{\text{H}0}$. We wish to construct solutions for which $\phi$ vanishes at infinity, so in our final result we should have $\phi_\infty = 0$. Using the symmetry~\eqref{eq:symmetrylambda}, we can impose this requirement at the last stage of our procedure: we perform a shooting on $\chi_{\text{H}0}$ using only the asymptotic behavior of $\chi$ and keeping $\phi_{\text{H}0}$ fixed at an arbitrarily chosen value; then, we evaluate
\begin{align}
    \phi_\infty = \phi(r_\infty) - \left[\frac{D}{r_\infty} + \frac{D \MBH}{r_\infty^2} - \frac{D(D^2 - 16 \MBH^2)}{12 r_\infty^3}\right] \, ,
    \label{eq:PhiInfEvaluation}
\end{align}
and apply the transformation~\eqref{eq:symmetrylambda} with $C = \phi_\infty$, obtaining a new value of $\lambda$ and the final value of $\phi_{\text{H}0}$.

From the numerical point of view this procedure is hard to carry out, as $\chi$ tends to diverge at infinity if it does not exactly match a BH solution. To overcome this difficulty, we first perform a preliminary search for the horizon value of $\chi$, either manually or semi-automatically, gradually pushing the outer boundary $r_\infty$ (and the divergence) to ever larger values as the interval containing $\chi_{\text{H}0}$ becomes narrower.
When we reach a situation in which the integration does not crash and the outer boundary radius is large enough that the asymptotic regime is reached, we proceed with an automatic shooting procedure with the bisection method to find the value of $\chi_{\text{H}0}$. Additionally, we work with a high accuracy, with the final solution being computed with $35$ significant figures for the quadratic case ($n = 2$) and $55$ significant figures for the quartic one ($n = 4$). Particularly in the quartic case, a lower number of significant figures either prevents the desired solution from being found or makes it impossible to reach a sufficiently large value of $r_\infty$ where the asymptotic behavior is recovered. The procedure is carried out using Wolfram \textsc{Mathematica}, which provides the functionality for very high-accuracy numerical integration. As for the relative accuracy in the final result of $\chi_{\text{H}0}$, the stopping threshold for the shooting procedure varies across the different sets, but is in the range of $10^{-20}$ and $10^{-50}$ for the quadratic and quartic cases, respectively.

Once the horizon parameters have been found, we reintegrate the equations \emph{both inside and outside} the horizon including the lapse $\alpha$. For the integration in the BH interior, we start at $\rH - \epsilon$ setting initial conditions as in Eq.~\eqref{eq:BCHorizonNumerical}, but with $\epsilon \to - \epsilon$. Given the residual gauge freedom in the metric, the lapse is defined up to a multiplicative constant $A$ (cf. Eq.~\eqref{eq:alphaEpxansionInf}), and we can set its horizon value to one. In practice we set $\alpha = 1$ both at $\rH \pm \epsilon$ in the two integrations. This amounts to introducing a discontinuity in $A$, which, however, does not affect our analysis, as we only use the lapse to evaluate the curvature invariants separately in the interior and in the exterior, and these do not depend on $A$. The integration in the interior is carried out up to the point $r = \rS$, where the denominator in the right-hand side of \cref{eq:DiffPhi} vanishes. As we shall argue in \cref{subsec:internal} by analyzing the curvature scalars, this point can be interpreted as a curvature singularity of these BH solutions.

Alongside these solutions, we also evaluate the spherically-symmetric BH solutions in EdGB gravity, in order to make comparisons. However, the procedure is much simpler and the shooting method is not required, as the field $\chi$ is not present~\cite{Corelli:2022phw,Corelli:2022pio}. In this case, we only set the horizon radius, together with $\phi_{\text{H}0}$ and $\lambda$. We then integrate the equations numerically, and estimate $\MBH$, $D$, and then $\phi_\infty$ at the outer boundary. Finally, we rescale $\phi$ and $\lambda$ according to the symmetry~\eqref{eq:symmetrylambda} in order to make the dilaton vanish asymptotically. We carry out the computation with at least $55$ significant figures as this allows us to better analyze the interior structure reducing the numerical noise close to the singularity. In very specific cases the numerical integration crashes, but increasing the accuracy to slightly more figures is sufficient to resolve the issue. 

Let us conclude this section by commenting on the relative difficulty in obtaining solutions in $f(\RR)$-dGB theory. In the quartic case, this task is particularly time-consuming due to the divergences at large $r$, which often cause crashes. Indeed, one must perform a large number of integrations with high precision, following the procedure manually. As a result, the entire process requires approximately $8\text{--}9\,\text{hrs}$ on a single core of a modern workstation. For comparison, constructing a BH configuration in EdGB (which does not require a shooting procedure) with the same numerical precision on the same machine takes only seconds.

\section{Results}
\label{sec:Results}
In this section, we present the main results of the numerical integration previously outlined, focusing on the domain of existence of BH solutions, on their internal structure, and discussing issues of well-posedness and hyperbolicity of the dynamical field equations inside the horizon. 

As anticipated in \cref{subsec:StaticBHSolutions}, we examine $f(\RR)$-dGB theories where the $f(\RR)$ term has the form~\eqref{eq:fRGeneral}, focusing on the quadratic and quartic cases, i.e., $n = 2, \, 4$, respectively. For concreteness, we set the dilaton coupling constant in \cref{eq:DilatonicCoupling} to $\gamma = 4$, although other values are expected to yield qualitatively similar results. 

\subsection{Behavior of the solutions}
\label{subsec:SolutionsBehavior}

Let us start our analysis by discussing the behavior of the solutions, first presenting the quadratic case and then the quartic one. 

\begin{figure}
    \centering
    \includegraphics[width=\columnwidth]{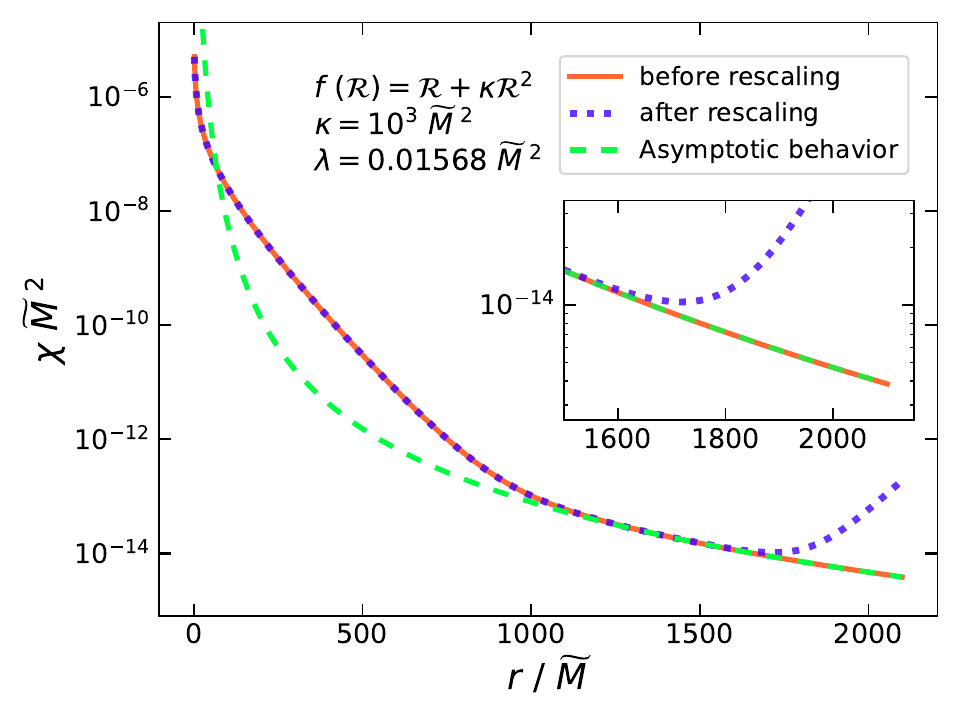} \\
    \caption{Exterior profile of the scalar field $\chi$ for a BH solution in the quadratic case ($n = 2$). The orange solid and the blue dotted lines denote the profile before and after the application of the symmetry transformation~\eqref{eq:symmetrylambda}, respectively. The green dashed line denotes the asymptotic behavior in Eq.~\eqref{eq:chiExpansionInfQuadratic}. While the application of the symmetry generally leaves the profile of $\chi$ unaltered, as expected, large discrepancies appear close to the outer boundary, probably due to the high sensitivity of these solutions to numerical errors.
    }
    \label{fig:SolutionProfilesQuadratic}
\end{figure}

In Fig.~\ref{fig:SolutionProfilesQuadratic} we present the profile of $\chi$ for a BH configuration in the quadratic case with $\kappa = 10^3 \,  \M^2$ (corresponding to $\ell \approx 31.6 \, \M$ in terms of the reference length scale $\ell$ introduced in \cref{sec:Theory&FE}). The solid orange curve is the result we obtained directly after the bisection procedure, before applying the transformation~\eqref{eq:symmetrylambda} to set $\phi_\infty = 0$; the blue dotted line denotes instead the profile of $\chi$ obtained after rescaling and re-integrating numerically the equations; the green dashed line is the asymptotic expansion we impose in our procedure, Eq.~\eqref{eq:chiExpansionInfQuadratic}. In the inset we show a zoom-in on the large $r$ region. We can see that the asymptotic behavior is reached accurately before the rescaling. The application of the symmetry transformation~\eqref{eq:symmetrylambda} introduces numerical errors, to which the solution is highly sensitive, requiring an extreme fine tuning of the horizon values. Such errors are nevertheless not relevant throughout most of the exterior domain up to the vicinity of the outer boundary, as indicated by the good agreement between the orange and blue curves.

\begin{figure}
    \centering
    \includegraphics[width=\columnwidth]{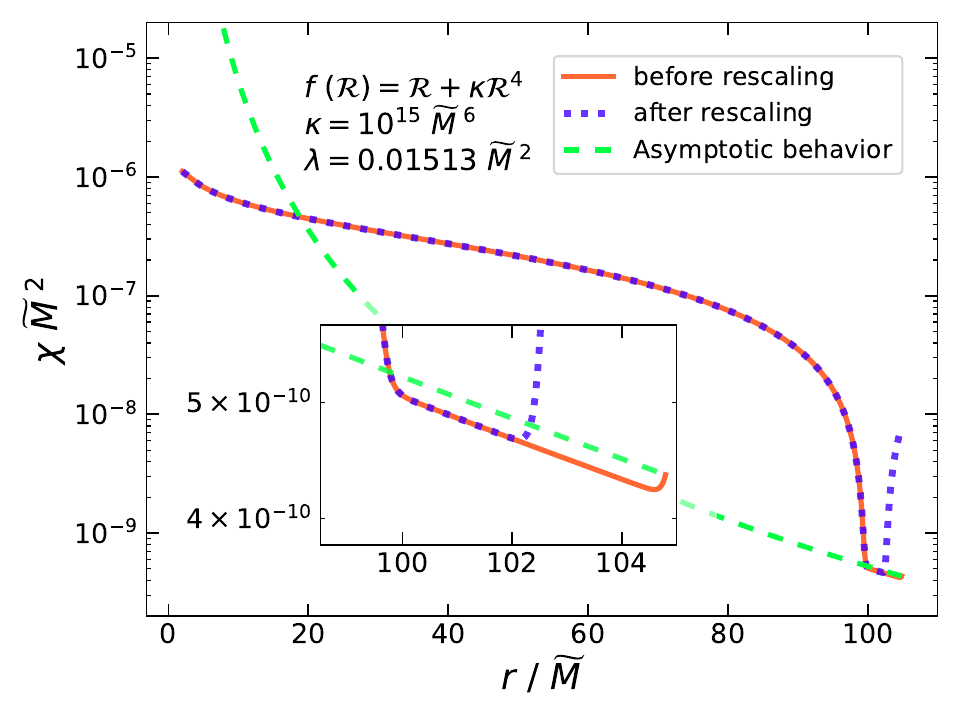} \\
    \includegraphics[width=\columnwidth]{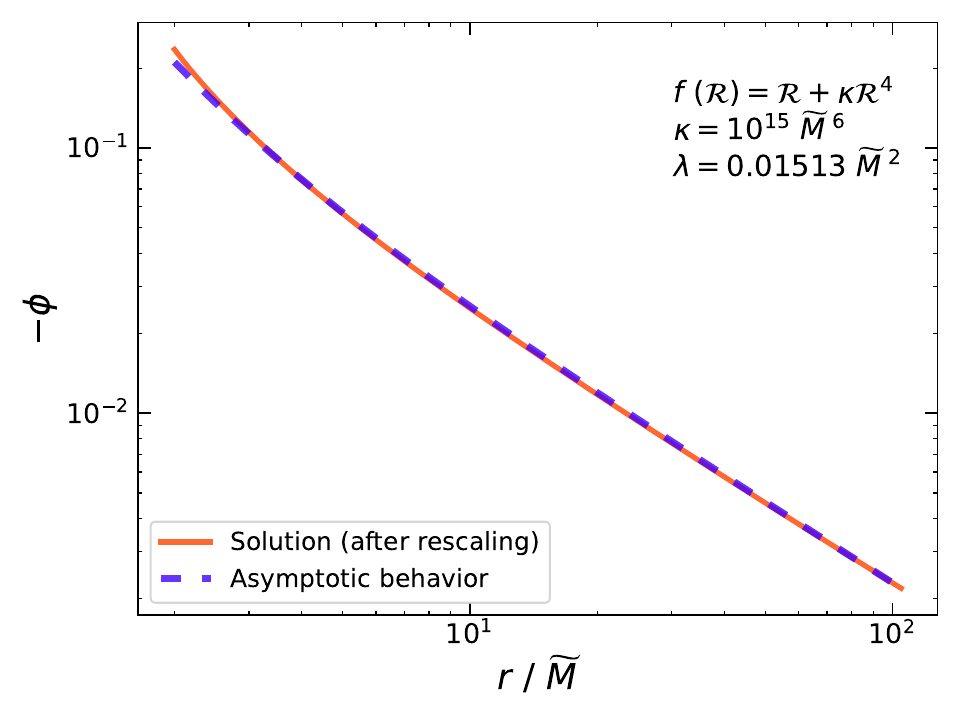}
    \caption{Exterior profiles of the scalar fields for a BH solution in the quartic case ($n = 4$). \textbf{Upper panel}: behavior of $\chi$. 
    Colors and conventions are the same as in the upper panel of Fig.~\ref{fig:SolutionProfilesQuadratic}, with the asymptotic behavior being now given by Eq.~\eqref{eq:chiExpansionInfQuartic}. The application of the symmetry transformation~\eqref{eq:symmetrylambda} introduces errors close to the outer boundary, as in the quadratic case. \textbf{Lower panel}: behavior of $\phi$ after applying the transformation~\eqref{eq:symmetrylambda} (orange solid line), compared against the asymptotic behavior in Eq.~\eqref{eq:phiExpansionInf} (blue dashed line).}
    \label{fig:SolutionProfilesQuartic}
\end{figure}

In the upper panel of Fig.~\ref{fig:SolutionProfilesQuartic}, instead, we show the profile of the $\chi$ field for a BH solution in the quartic case with $\kappa = 10^{15} \, \M^6$, using the same conventions as in Fig.~\ref{fig:SolutionProfilesQuadratic} (with the asymptotic behavior now given by \cref{eq:chiExpansionInfQuartic}).
Note that its unexpectedly large value actually corresponds to a more reasonable length scale $\ell \approx 316 \, \M$, owing to the fact that, in the quartic case, $\kappa\equiv \ell^6$. The value of the coupling was chosen as it allowed us to evaluate the solutions in reasonable time. For smaller values, both the integration time and the difficulty in tuning the correct boundary condition $\chi_{\text{H}0}$ in the shooting method described in \cref{subsubsec:NumericalIntegration} grew significantly large.

Close to the outer boundary, $\chi$ shows a sharp increase and crosses the green dotted line. This is related to the fact that $\chi$ tends to diverge at large radii even for very small discrepancies from the exact BH solution. Our algorithm pushes the crossing between the blue and green curve to increasingly large values of $r$, postponing the divergence and allowing $\chi$ to attain its asymptotic regime, within a certain degree of accuracy. 
The same issue does not to happen for $\phi$, which is shown in the lower panel of the same figure. For some sample solutions, we also computed the mass and scalar charge before the onset of the sharp increase, obtaining relative discrepancies with the values at the outer boundary that are $\lesssim 10^{-5}$ for the mass and $\lesssim 10^{-3}$ for the scalar charge. Since they are computed from $\phi$ and $\zeta$ (see \cref{eq:MDAsymptoticEstimate}), this provides a further check that their profiles are not affected by any diverging behavior. Also in this quartic case, after performing the transformation~\eqref{eq:symmetrylambda}, the $\chi$ solution exhibits a rapid increase near the outer boundary, again likely due to numerical errors. Nevertheless, these errors are again not relevant throughout most of the domain and, additionally, as blue and red lines match well before that point, confirming the reliability of the result.

\subsection{Mass-radius diagram and minimum mass}
\label{subsec:ExistenceDomain}
In standard GR, static neutrally-charged BHs feature a linear relation between the mass and the event horizon radius, and BHs with arbitrarily small radii and masses exist within the theory's spectrum. By contrast, in EdGB gravity, the coupling to the dilaton introduces an additional length scale, set by the coupling constant $\lambda$, which gives rise to more intricate features~\cite{Torii:1996yi,Corelli:2022phw,Corelli:2022pio}. First, static BHs in EdGB cannot exist below a minimum mass and radius; instead, there exists a minimum mass configuration, which is distinguished from a minimum $\rH$ one. The minimum-mass solution divides the solution spectrum into two branches. The one characterized by the larger radii is dynamically stable. For very large values of $\rH^2/\lambda$, this is the only branch that survives, and its solutions asymptotically (for $\rH/\sqrt{\lambda}, \, \MBH/\sqrt{\lambda} \gg 1$) approach the GR relation $\rH = 2\MBH$. The other branch, associated with smaller radii, is unstable and terminates at a configuration that is singular at the event horizon (see, e.g., Refs.~\cite{Kanti:1995vq,Torii:1996yi,Kanti:1997br,Torii:1998gm,Ripley:2019aqj,Corelli:2022phw,Corelli:2022pio}). 

The existence of a minimum-mass state in the spectrum also introduces conceptual difficulties in describing BH evaporation within EdGB gravity~\cite{Corelli:2022phw,Corelli:2022pio}. Indeed, the temperature and graybody factors of these solutions remain nonvanishing, implying that Hawking radiation~\cite{Hawking:1975vcx} should persist even when the system reaches the minimum-mass configuration. However, since the spectrum lacks lower-energy states to which the system can decay, a complete description of the evaporation process is precluded. This issue is further compounded by the appearance of a naked elliptic region in the final stages of evaporation, which renders the theory nonpredictive in this regime~\cite{Corelli:2022phw,Corelli:2022pio}.

\begin{figure}[!ht]
    \centering
\includegraphics[width=\linewidth]{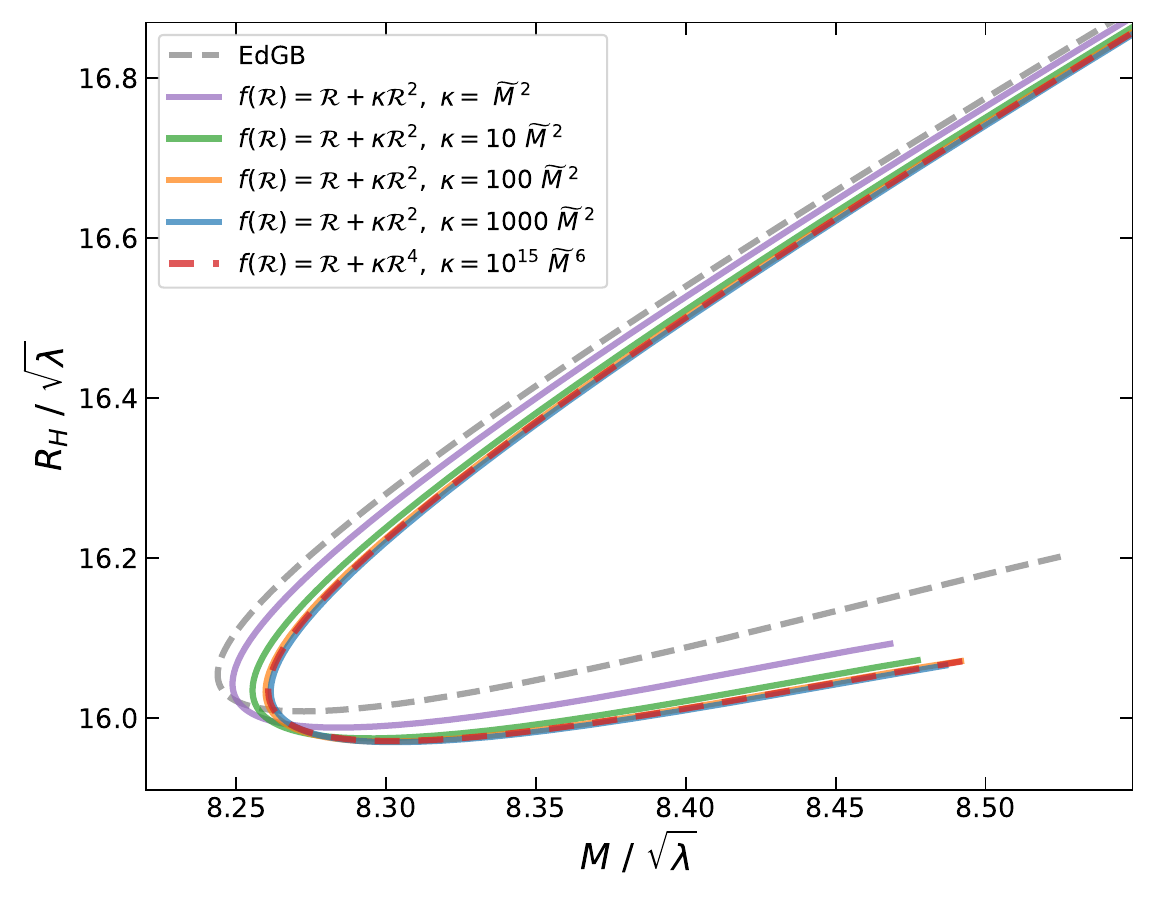}
    \caption{Event horizon radius as a function of BH mass, both normalized by the square root of the dilaton coupling constant $\lambda$, for different theories. The dashed gray line corresponds to EdGB theory with the exponential coupling~\eqref{eq:DilatonicCoupling}. The other curves correspond, instead, to the $f(\RR)$-dGB theory, with $f(\RR) = \RR + \kappa \RR^n$. Solid lines correspond to the quadratic case ($n = 2$) for different values of $\kappa/\M^2=\left(1,10,100,1000\right)$, corresponding to $\ell/\M \approx \left(1, 3.16, 10, 31.6\right)$ in terms of the reference length scale $\ell$ (violet, green, orange and blue, respectively). The dashed red line corresponds to the quartic case ($n = 4$) with $\kappa = 10^{15} \, \M^6$, corresponding to $\ell \approx 316 \, \M$.}
    \label{fig:DomainExistence}
\end{figure}

In what follows, we extend this analysis to $f(\RR)$-dGB theory, also assessing possible differences with the standard EdGB picture outlined above due to higher-curvature terms. 

For the quadratic case, we examined four representative regimes: $\kappa = \left(1, 10, 100, 1000\right) \, \M^2$, corresponding to the reference length scale $\ell \approx \left(1, 3.16, 10, 31.6\right)\, \M$.
In the quartic case, instead, we present results for a single representative value, $\kappa = 10^{15}\,\M^6$ (corresponding to $\ell \approx 316\,\M$), which serves as a paradigmatic example. This choice is also motivated by the substantially higher computational cost discussed in the previous section.

To accelerate the construction of the mass-radius diagrams, we limited our sampling to roughly $20$ solutions per theory and value of $\kappa$. For each solution, we extracted the corresponding mass $\MBH$ and rescaled value of $\lambda$ following the procedure outlined in \cref{subsubsec:NumericalIntegration}. This provided sufficiently smooth curves, which were further refined by cubic spline interpolation to produce the curves shown in \cref{fig:DomainExistence}. The dashed gray curve corresponds to purely EdGB solutions, while solid lines and the dashed red line to $f(\RR)$-EdGB ones in the quadratic and quartic case, respectively. Several features are worth emphasizing. 

First, although BH solutions in $f(\RR)$-dGB gravity differ quantitatively from EdGB ones, they retain 
qualitatively the same nonperturbative features: (i) two distinct branches of solutions separated by a minimum-mass configuration; (ii) the existence of a minimum-radius configuration; (iii) a lower branch terminating at a solution that is singular at the horizon. 
As in EdGB theory, the existence of a BH configuration with a minimum mass may raise a conundrum regarding Hawking evaporation~\cite{Torii:1996yi, Alexeyev:2002tg, Corelli:2022pio, Corelli:2022phw}. Indeed, if such BHs lose mass through Hawking radiation, they would eventually enter a regime in which no static BH solutions exist, leaving their subsequent evolution unclear. To determine whether this scenario can actually occur, one should verify whether Hawking evaporation persists at the minimum mass by computing the corresponding temperature and graybody factors. Although we expect this to be the case also in $f(\RR)$-dGB theory, a specific analysis is left for future work.

Second, in the quadratic case, the mass-radius diagrams tend to approach the same profile as $\kappa$ increases. For instance, the curves for $\kappa = 100 \, \M^2$ and $\kappa = 1000 \, \M^2$ are nearly coincident, whereas those for $\kappa = \M^2$ and $\kappa = 10 \, \M^2$ display a larger separation. Remarkably, this behavior seems to also extend across different powers of the higher-curvature terms: the quadratic case with $\kappa = 1000 \, \M^2$ nearly overlaps the quartic case with $\kappa = 10^{15}\, \M^6$, despite arising from distinct theories and vastly different couplings. This suggests the possible existence of a universal behavior for $n\geq2$ in the large-coupling limit. We further discuss this observation and propose a tentative explanation in the following section. 

Finally, in the quadratic theory, decreasing $\kappa$ shifts the diagram toward the EdGB one; this is expected, despite the subtitles in the limit $\kappa \to 0$ highlighted below Eq.~\eqref{eq:Diffalpha}-\eqref{eq:DiffChi}. 

\subsection{Internal structure of the solutions}
\label{subsec:internal}

In EdGB gravity, there is strong evidence for the presence of an inner high-curvature region, as indicated by a rapid increase of the curvature scalars (see, e.g., Refs.~\cite{Alexeev:1996vs,Sotiriou:2013qea,Sotiriou:2014pfa}), while the theory also features a breakdown of hyperbolicity~\cite{Ripley:2019aqj,Ripley:2019irj,East:2021bqk,Corelli:2022phw,Corelli:2022pio,R:2022hlf}. The striking similarity between the exterior properties of $f(\RR)$-dBG and EdGB BH solutions, noticed in the previous section, naturally prompts the question of to what extent higher-curvature operators modify the interior structure. Here we focus on the internal structure of the $f(\RR)$-dGB solution and the role of the singularity. The hyperbolicity and well-posedness issues will be considered in the next section.

\begin{figure}
    \centering
    \subfigure{\includegraphics[width=\linewidth]{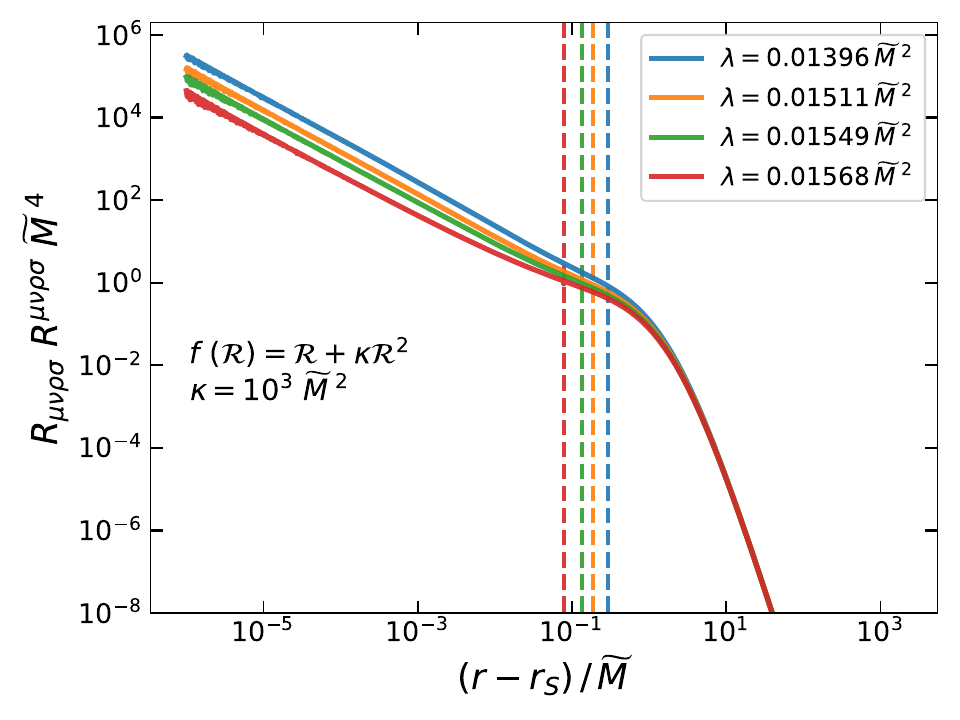}}
    \subfigure{\includegraphics[width=\linewidth]{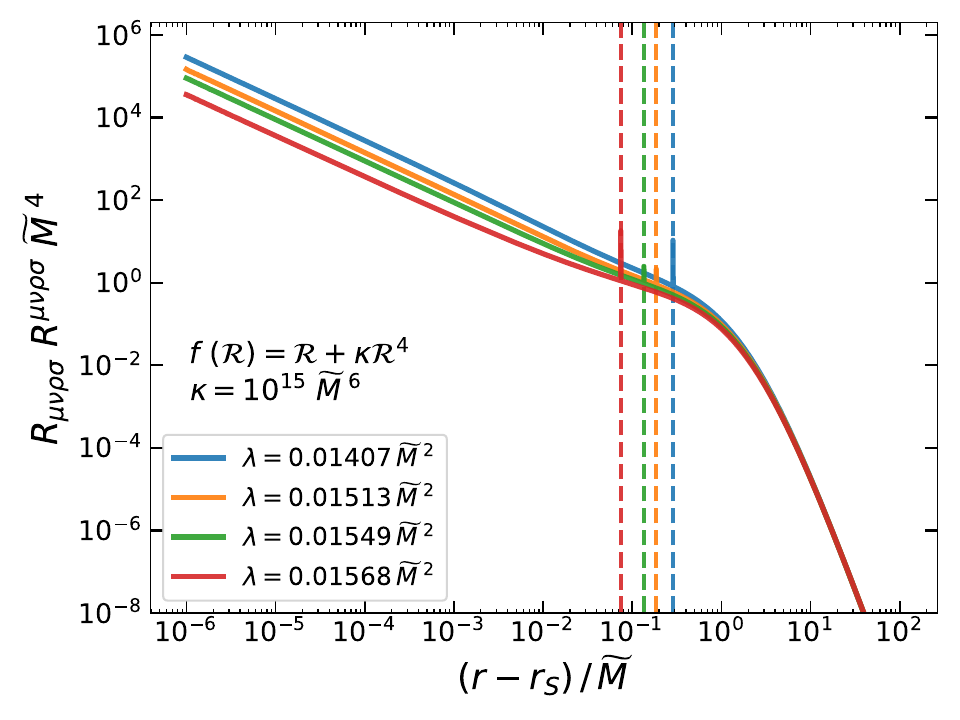}}
    
    \caption{Kretschmann scalar (solid curves) as a function of the distance from $r_\text{S}$ for representative BH solutions of $f(\RR)$-dGB gravity, with particular values of $\kappa$ and different values of the dilaton coupling constant $\lambda$. The top (bottom) panel shows the quadratic (quartic) $f(\RR)$–dGB case. In both figures, the vertical dashed lines mark the position of the horizons of each solution and are color–coded consistently with the corresponding solid curves.}
    \label{fig:kretschmannscalar}
\end{figure}

\begin{figure}
    \centering
    \subfigure{\includegraphics[width=\linewidth]{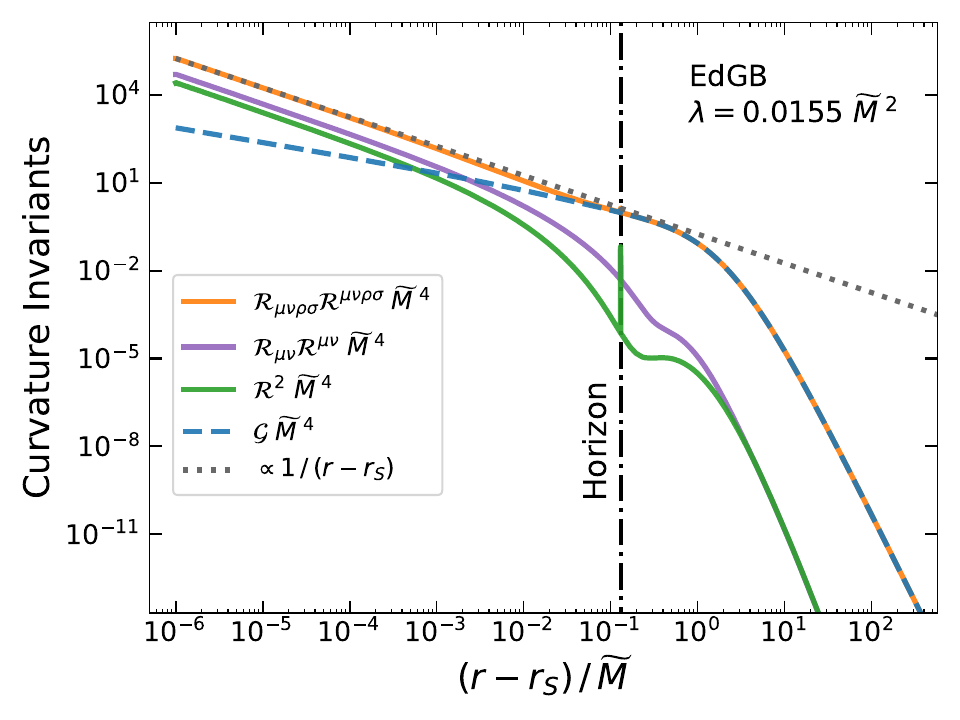}}
    \subfigure{\includegraphics[width=\linewidth]{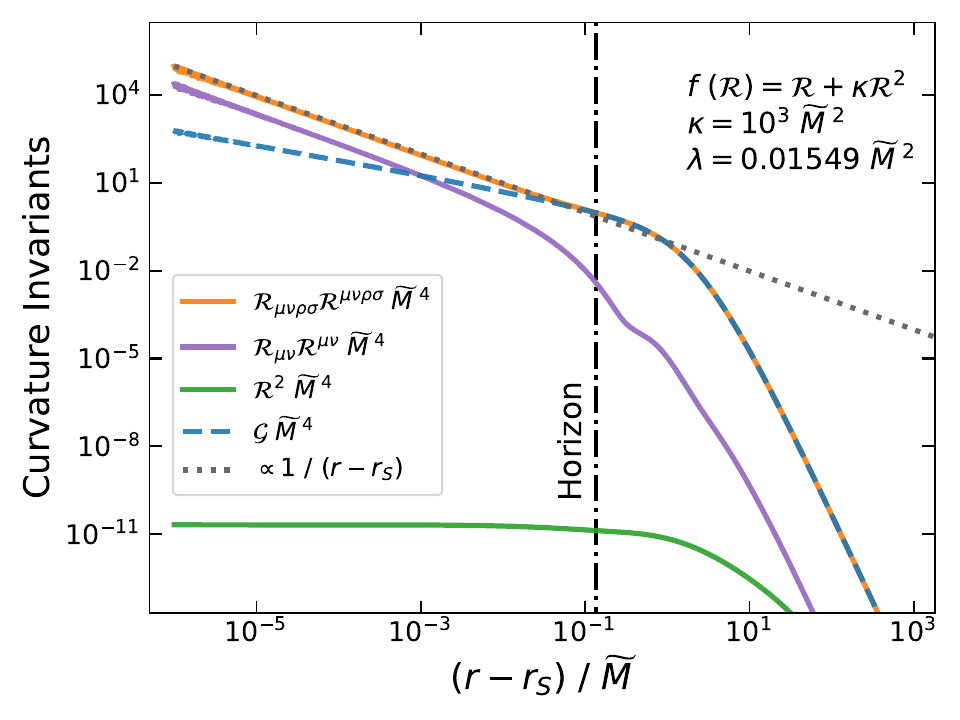}}
    \subfigure{\includegraphics[width=\linewidth]{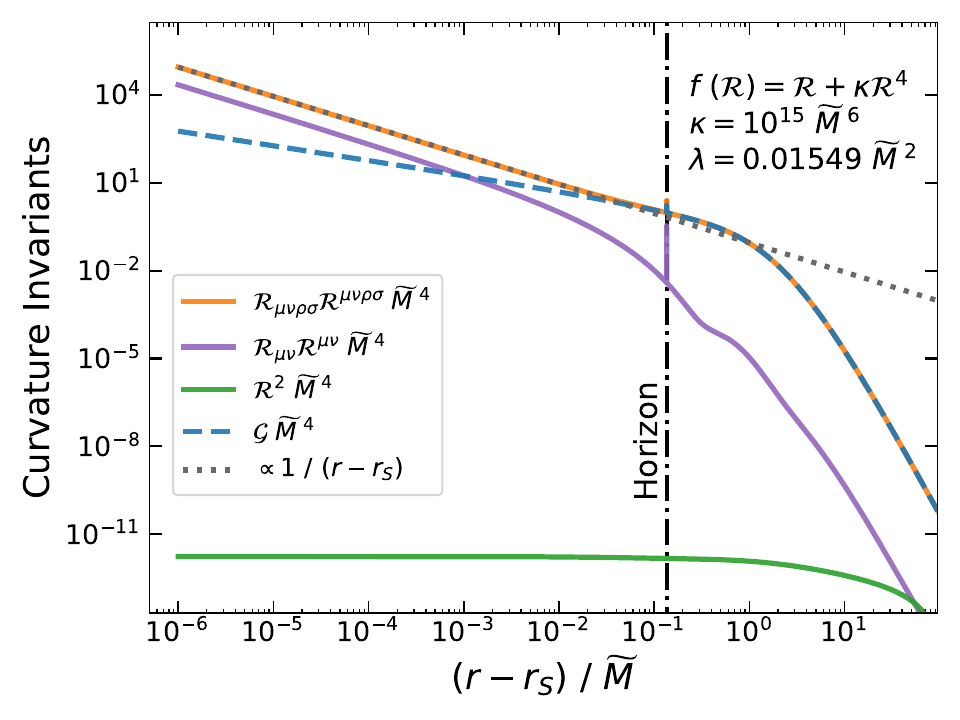}}
    
    \caption{Behavior of the curvature invariants as a function of $r-\rS$ in EdGB, quadratic and quartic $f(\RR)$-dGB theories (top, middle and bottom panels, respectively), shown for representative values of the coupling constants $\kappa$ and $\lambda$. The gray dotted line marks a reference scaling $\sim (r-\rS)^{-1}$ for the Kretschmann scalar in the interior region. In all panels, the vertical dot-dashed line indicates the location of the event horizon.}
    \label{fig:curvatureinvariants}
\end{figure}

In our investigations, we found that, despite being regular at the horizon, as imposed by the boundary conditions (see \cref{subsubsec:BoundaryConditions}), all $f(\RR)$-dGB BH solutions encountered a point ---denoted by $r_\text{S}$ --- at which the integration breaks down and terminates. We numerically verified that this point corresponds to a zero of the denominator of the right-hand side of \cref{eq:DiffPhi}, preventing the integration from being extended beyond. To assess whether $r_\text{S}$ corresponds to a genuine curvature singularity, we computed the Kretschmann scalar for representative solutions with different values of the dilaton coupling constant $\lambda$. The results are reported in \cref{fig:kretschmannscalar}, where the Kretschmann scalar is plotted as a function of the radial distance from $r_\text{S}$. In the quadratic case, we show solutions with $\kappa = 10^3 \, \M^2$ as representative examples. The computation was carried out by considering separately the solution in the interior and exterior. This leads to the presence of the spikes at the horizon (mostly noticeable in the lower panel of \cref{fig:kretschmannscalar}), which is located at the boundary of the two domains of integration considered. In both quadratic and quartic cases, the Kretschmann scalar grows by several orders of magnitude as $r \to r_\text{S}$. This indicates that the region around $r_\text{S}$ is characterized by an extremely large spacetime curvature, consistent with the interpretation of $r_\text{S}$ being a curvature singularity. As illustrated in \cref{fig:curvatureinvariants}, this growth follows an approximate $(r-\rS)^{-1}$ scaling, milder than the $r^{-6}$ divergence of the Schwarzschild solution but qualitatively similar to EdGB BHs. This could be interpreted as tentative evidence supporting the idea that higher-curvature terms in the gravitational action might help, at least partially, to mitigate the classical BH singularity problem—although an infinite tower of such corrections may ultimately be required to fully resolve the issue~\cite{Bueno:2024dgm,Bueno:2025zaj}. 

\begin{figure}
    \centering
    \includegraphics[width=\linewidth]{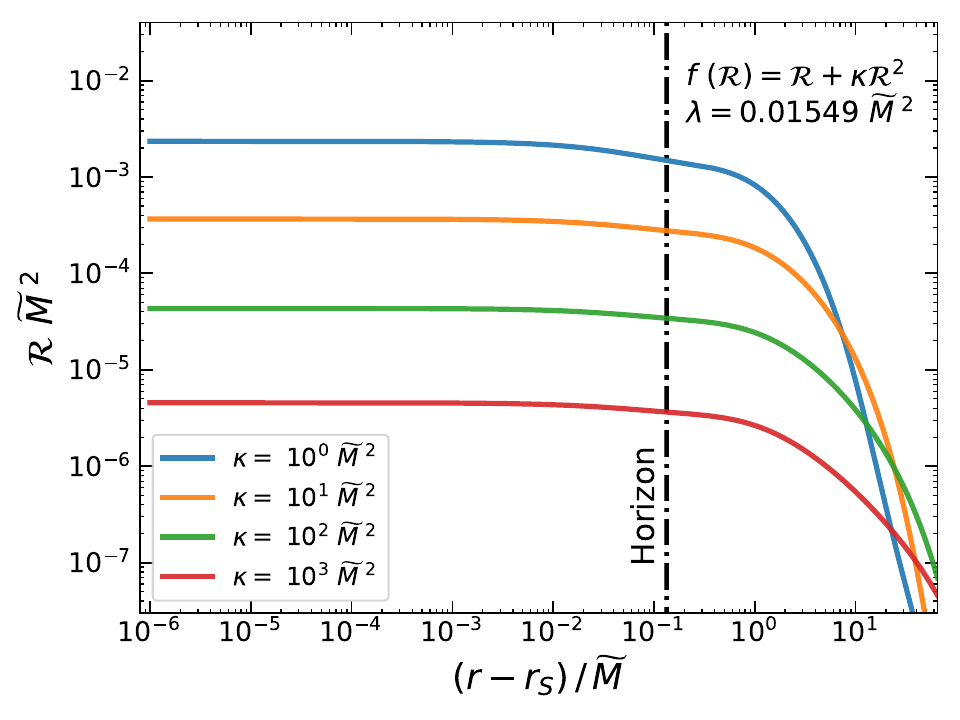}
    \caption{Ricci scalar, as a function of $r-\rS$, for the quadratic $f(\RR)$-dGB theory, at fixed value of the dilatonic coupling constant $\lambda$, but at varying $\kappa$. The vertical black lines (which are practically all overlapping on each other) mark the position of the horizons of each solution. The continuity of $\RR$ at the horizon confirms the reliability of the algorithm in constructing the interior solution.}
    \label{fig:RicciVaryingkappa}
\end{figure}

As shown in \cref{fig:curvatureinvariants}, also other curvature scalars exhibit a rapid increase as $r \to \rS$,  both in EdGB and $f(\RR)$-dGB solutions.
A striking difference, however, emerges in the behavior of the square of Ricci scalar in the interior. In EdGB theory, its magnitude is comparable to the other curvature invariants (top panel in \cref{fig:curvatureinvariants}), whereas in $f(\RR)$-dGB theory it is suppressed by $\sim 15$ orders of magnitude (for the examples shown in the middle and lower panels in \cref{fig:curvatureinvariants}). It also exhibits an approximately constant trend in the interior and remains regular at $\rS$, in contrast to the EdGB case. Crucially, this behavior appears to be qualitatively the same across different coupling regimes, i.e., for various values of $\kappa$. Moreover, larger values of $\kappa$ lead to a stronger suppression of $\RR$ (see \cref{fig:RicciVaryingkappa} for the quadratic case). These differences highlight a profound structural difference in the interior of $f(\RR)$-dGB BHs compared to EdGB ones. However, since the Gauss-Bonnet term diverges as $r \to \rS$, regardless of the value of $\kappa$, its contribution will always dominate over the Ricci term (and powers thereof) at $\rS$. Therefore, although the interior structure is quantitatively different, this \emph{dynamical} mechanism (we recall that the Ricci scalar is an independent degree of freedom in this theory) makes the singularity structure of $f(\RR)$-dGB solutions qualitatively the same as that of the EdGB BHs (see also the discussion in the next section).

\begin{figure}
    \centering
    \subfigure{\includegraphics[width=\linewidth]{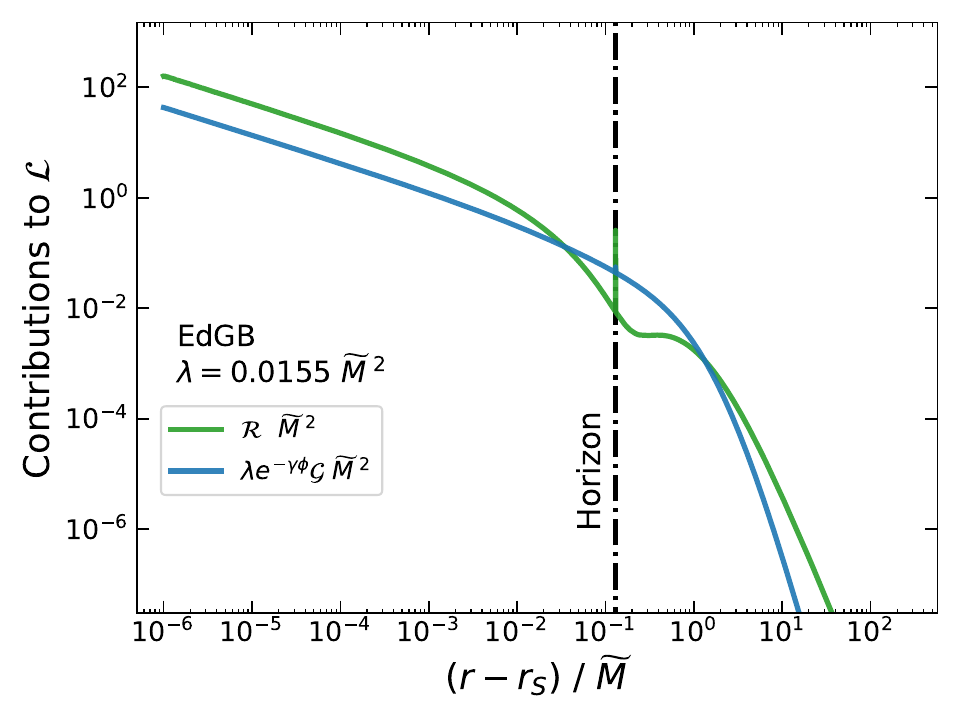}}
    \subfigure{\includegraphics[width=\linewidth]{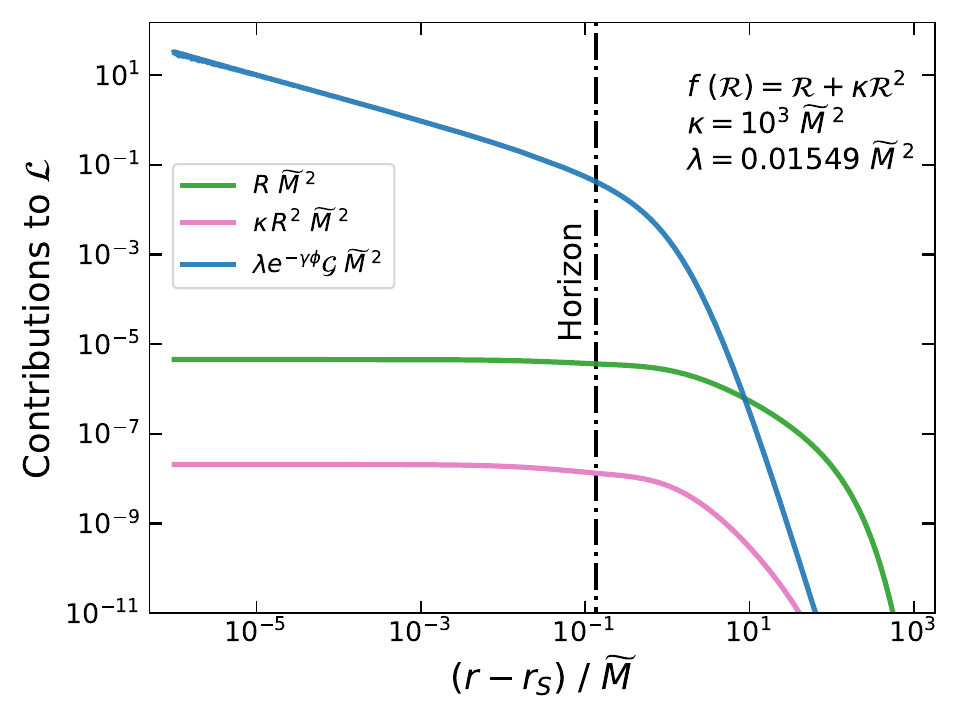}}
    \subfigure{\includegraphics[width=\linewidth]{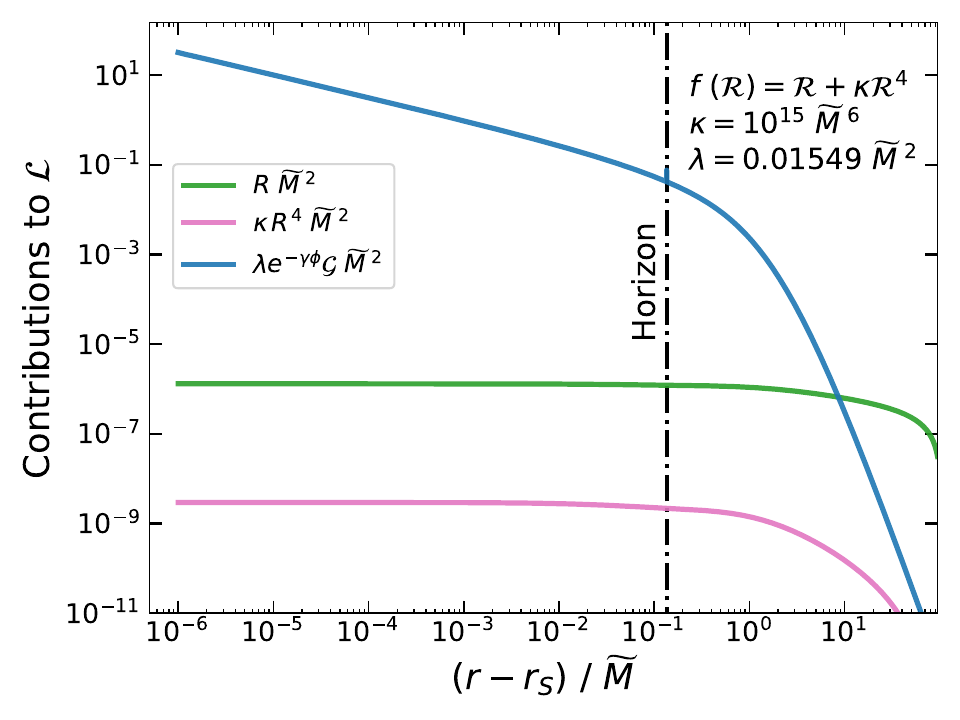}}
    \caption{Comparison of the different contributions to the Lagrangian density in~\eqref{eq:Action_JPS} as a function of the radial distance from the central singularity $r_\text{S}$. The top panel shows the EdGB case (with higher-curvature terms coming from the $f(\RR)$ switched off), where the Ricci scalar contribution is compared with the dGB term (green and blue solid lines, respectively). The middle and bottom panels display the $f(\RR)$–dGB cases, quadratic and quartic, respectively, including the higher-curvature operators (pink solid line) in the comparison. In all figures, the vertical dot-dashed line represents the location of the event horizon.}
    \label{fig:LagrangianContributions}
\end{figure}

Given the radically different structure of the field equations and the structural difference of the interior, it is natural to ask why the exterior phenomenology of $f(\RR)$-dGB BHs so closely resembles that of EdGB. To understand this issue, the evaluation of the different contributions appearing in the Lagrangian of \cref{eq:Action_JPS} turns out to be useful. A detailed comparison between them is reported in \cref{fig:LagrangianContributions}.  As shown in this figure, the dGB term consistently dominates over the quadratic, and even quartic, curvature corrections, both inside and outside the horizon, for the values of $\kappa$ considered. This explains why the global, nonperturbative features of EdGB remain qualitatively valid also in $f(\RR)$-dGB gravity, despite the higher-curvature operators modifying quantitatively the interior structure of the solutions. We shall return to this point in the next section when discussing the hyperbolicity and well-posedness. 
Finally, a comparison between the pink curves in \cref{fig:LagrangianContributions} (corresponding to the $\kappa \RR^2$ and $\kappa \RR^4$ contributions to the action) reveals that the higher the power of the curvature correction, i.e., $n$ in \cref{eq:fRGeneral}, the \emph{more suppressed} the contribution. This may explain the rather unexpected similarity observed in the mass-radius diagrams of the quadratic and quartic cases, in the strong coupling regime, which could hint to a universal behavior for $f(\RR)$-dGB theories with $n \geq 2$.

\subsection{Hyperbolicity and well-posedness}
\label{subsec:HyperbolicityWellPosedness}

We now proceed to study the local well-posedness of $f(\RR)$-dGB theory, comparing it against the EdGB case. For this purpose we should consider the non-stationary case (still restricted to spherical symmetry), obtain the system of evolution equations, and compute their principal symbol. We will do this using the same formalism as in Refs.~\cite{Ripley:2019aqj, Corelli:2022phw} which applies to a first-order system (see also~\cite{Sarbach:2012pr, Whitham:1999lnw, Reula:1998ty} for standard references on hyperbolicity). Finally, by evaluating the principal symbol on the static solutions discussed in the previous section, we can check the region where the system is strongly hyperbolic.

To obtain the system of evolution equations, we shall first allow for a time dependence in the metric functions and the fields. In other words, we now consider $\alpha$, $\zeta$, $\phi$ and $\chi$ as function of $(t, r)$. We then introduce auxiliary variables for the scalar fields, representing their radial derivatives and their momenta (in the following, derivatives with respect to $t$ and $r$ are written explicitly to avoid ambiguity):
\begin{align}
    Q = \rder \phi, \quad & \quad P = \frac{1}{\alpha} \tder \phi - \zeta Q \, , \notag \\
    \Theta = \rder \chi, \quad & \quad \Pi = \frac{1}{\alpha} \tder \chi - \zeta \Theta \, .
    \label{eq:AuxiliaryEvolutionVariables}
\end{align}
From these definitions we can then extract the evolution equations for $\phi$ and $Q$, as
\begin{align}
    \tder \phi &= \alpha P + \alpha \zeta Q \, , \label{eq:phitder} \\
    \tder Q &= \tder \rder \phi = \rder \left( \alpha P + \alpha \zeta Q \right) \, , \label{eq:Qtder}
\end{align}
and the same for $\chi$ and $\Theta$. Note that the equations for $\tder \phi$ and $\tder \chi$ are redundant, as the profile of these fields is already encoded in the auxiliary variables $Q$ and $\Theta$.

Next, we replace the auxiliary variables in the field equations and perform algebraic manipulations that allow us to remove the second derivative terms such as $\rder^2 \zeta$, $\tder \rder\zeta$, $\rder^2 \alpha$ and $\tder \rder \alpha$. After some other manipulation we can obtain evolution equations for $\zeta$, $P$ and $\Pi$ as well as constraints for the metric functions $\alpha$ and $\zeta$.

With the system of equations at hand we are now in the position of computing the principal symbol. The details and the structure of the system in EdGB and $f(\RR)$-dGB differ, so here we will describe the basic strategy, and later we will discuss the two cases separately.

Given a set of $N$ variables $v^I$ and a system of $N$ first order equations for them, $E_{v^I}$, we can define the principal symbol as
\begin{equation}
    \mathcal{P}_{IJ}(\eta_\mu) = \frac{\delta E_{v^I}}{\delta \partial_\mu v^J} \eta_\mu\, ,
    \label{eq:PrincipalSymbolDef}
\end{equation}
where $\eta^\mu$ is a 4-vector. Strong hyperbolicity can then be assessed by verifying the existence of a complete set of real $\eta_\mu$ that solve the characteristic equation
\begin{equation}
    \det \mathcal{P}(\eta_\mu) = 0 \, .
    \label{eq:Characteristic}
\end{equation}

In order to check where this condition holds, we now need to consider our specific systems.

\subsubsection{Principal symbol in EdGB theory}

In EdGB gravity the $\chi$ field is not present and the set of our variables is $v^I = \left( \phi, Q, P, \alpha, \zeta \right)$. The set of equations is instead given by Eq.~\eqref{eq:phitder},~\eqref{eq:Qtder} written as
\begin{align}
    \tder \phi &- \left (\alpha P + \alpha \zeta Q \right) = 0 \, , \label{eq:phitderImplicit} \\
    \tder Q &- \rder \left( \alpha P + \alpha \zeta Q \right) = 0 \, , \label{eq:QtderImplicit}
\end{align}
the evolution equation for $P$ written in implicit form, and the two constraints for $\alpha$ and $\zeta$ written in implicit and coupled form. With this we compute the principal symbol and its determinant, obtaining that it is proportional to $\eta_t \eta_r^2$. This derives from the redundancy of the evolution equation for $\phi$, and from the fact that $\alpha$ and $\zeta$ are constrained degrees of freedom. The remaining part is an homogeneous polynomial of degree $2$ in $\eta_t$ and $\eta_r$ which, factoring $\eta_r^2$ out gives
\begin{equation}
    \det \mathcal P \propto \mathfrak{a} \left(\frac{\eta_t}{\eta_r} \right)^2 + \mathfrak{b} \left(\frac{\eta_t}{\eta_r} \right) +  \mathfrak{c} \, ,
    \label{eq:CharacteristicPolynomialEdGB}
\end{equation}
where $\mathfrak{a}$, $\mathfrak{b}$ and $\mathfrak{c}$ are lengthy expressions that contain the fields, the metric functions and their derivatives.
Therefore, the existence of a complete set of real $\eta_\mu$, and hence strong hyperbolicity, is guaranteed by the positivity of the discriminant
\begin{equation}
    \Delta = \mathfrak{b}^2 - 4 \mathfrak{a} \mathfrak{c} \, .
    \label{eq:EdGBDiscriminant}
\end{equation}
In practice, to simplify the computation, we take advantage of the fact that we are interested in assessing the hyperbolicity on a static configuration, and in Eq.~\eqref{eq:CharacteristicPolynomialEdGB} we require that all the variables depend only on the radial coordinate. 
Note that doing so at this stage does not affect the hyperbolicity of the system, as the computation of the principal symbol, which involves derivatives with respect to $\partial_\mu v^J$, has already been carried out.

\subsubsection{Principal symbol in $f(\RR)$-dGB theory}

Let us now move to $f(\RR)$-dGB theory. In this case we also have the scalar field $\chi$ so that the set of variables is $v^I = \left( \phi, Q, \chi, \Theta, P, \Pi, \alpha, \zeta \right)$. The set of equations is instead composed by the four equations for $\phi$, $Q$, $\chi$ and $\Theta$ written in the same form as Eqs.~\eqref{eq:phitderImplicit}-\eqref{eq:QtderImplicit}, the evolution equations for $P$ and $\Pi$ written in implicit and coupled form, and the constraints for $\alpha$ and $\zeta$ written in implicit and coupled form. Using Eq.~\eqref{eq:PrincipalSymbolDef} we compute the principal symbol, which is lengthy and extremely hard to manipulate even with a computer algebra software. We therefore impose staticity directly on $\mathcal{P}$, before computing the determinant, in order to simplify the computation. In particular here we impose that all the fields and the metric functions depend only on the radial coordinate, and we substitute the expression of $f(\RR)$ here (with the appropriate exponent $n$) to allow for further simplification. Furthermore, we also remove the occurrences of $\rder \alpha$ and $\rder \zeta$ using the constraints for $\alpha$ and $\zeta$, also restricted to the static case and to the choice of $f(\RR)$ under consideration. As discussed earlier this does not affect the result, as we are only interested in assessing the hyperbolicity in static BH spacetimes, and the discriminant of the principal symbol has to be evaluated on-shell.
Then, we compute the determinant, $\det \mathcal P$. In this case it is proportional to $\eta_t^2 \eta_r^2$, as now there are two redundant degrees of freedom ($\phi$ and $\chi$), and $\alpha$ and $\zeta$ are constrained. The remaining term is now a homogeneous polynomial of degree $4$, which can be written as
\begin{equation}
    \det \mathcal P \propto \mathfrak{a} \left(\frac{\eta_t}{\eta_r} \right)^4 + \mathfrak{b} \left(\frac{\eta_t}{\eta_r} \right)^3 + \mathfrak{c} \left(\frac{\eta_t}{\eta_r} \right)^2 + \mathfrak{d} \left(\frac{\eta_t}{\eta_r} \right) + \mathfrak{e} \, ,
    \label{eq:CharacteristicPolynomialfRdGB}
\end{equation}
after factoring $\eta_r^4$ out. The fact that Eq.~\eqref{eq:CharacteristicPolynomialfRdGB} is a polynomial of fourth degree reflects the fact that now there are two scalar fields.

Strong hyperbolicity is guaranteed if the roots of~\eqref{eq:CharacteristicPolynomialfRdGB} are real and distinct. This happens if all the following conditions hold~\cite{Rees01021922}:
\begin{align}
    \Delta &> 0 \, , \notag \\
	64 \mathfrak{a}^3 \mathfrak{e} - 16 \mathfrak{a}^2 \mathfrak{c}^2 + 16 \mathfrak{a} \mathfrak{b}^2 \mathfrak{c} - 16 \mathfrak{a}^2 \mathfrak{b d} - 3 \mathfrak{b}^4 &< 0 \, , \notag \\
	8 \mathfrak{a c} - 3 \mathfrak{b}^2 &< 0 \, ,
    \label{eq:HyperbolictyConditionfR}
\end{align}
where $\Delta$ is the discriminant of the polynomial. Away from these conditions it is still possible to have $4$ real roots, but only if the discriminant $\Delta$ vanishes. Therefore, to check the strong hyperbolicity we identified the point $\rE$ where at least one of the conditions in Eq.~\eqref{eq:HyperbolictyConditionfR} breaks down by means of a bisection strategy in the range $\rS \lesssim r \lesssim \rH$; then we sampled $100$ points in the range $\rS \lesssim r \lesssim \rE$ and $100$ in $\rE \lesssim r \lesssim \rH$, and we checked that $\Delta$ is negative in the interior of $\rE$ and positive in the exterior of it. In this way we could ensure that the system is strongly hyperbolic only where the conditions~\eqref{eq:HyperbolictyConditionfR} are satisfied and there are 4 distinct real solutions for $\eta_t / \eta_r$. 

\subsubsection{Breakdown of hyperbolicity on BH configurations}

\begin{figure}
    \centering
    \includegraphics[width=\columnwidth]{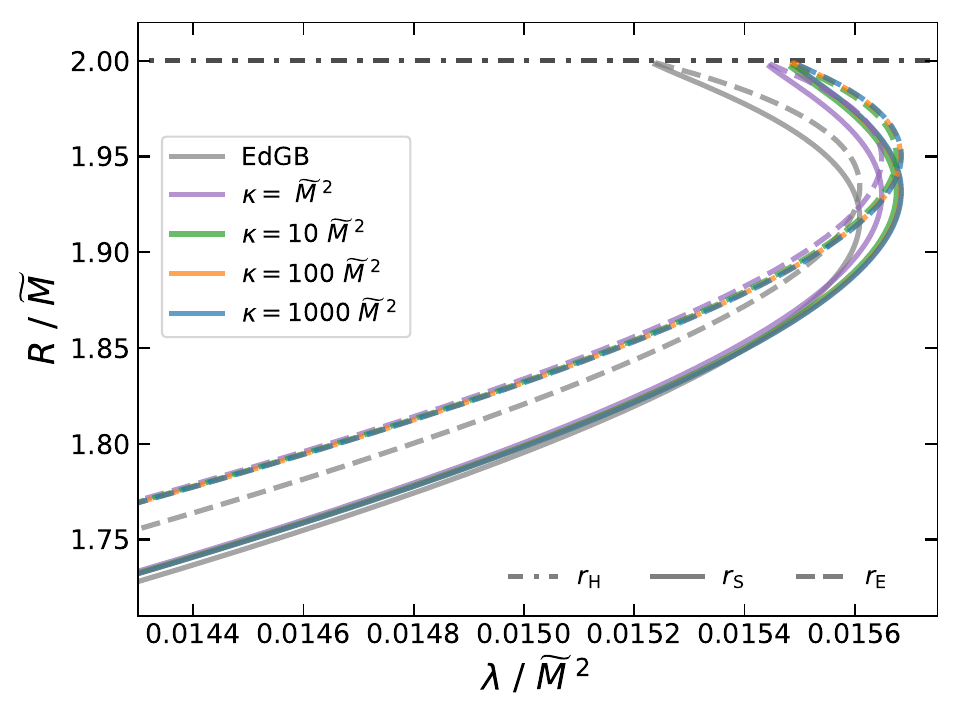} \\
    \includegraphics[width=\columnwidth]{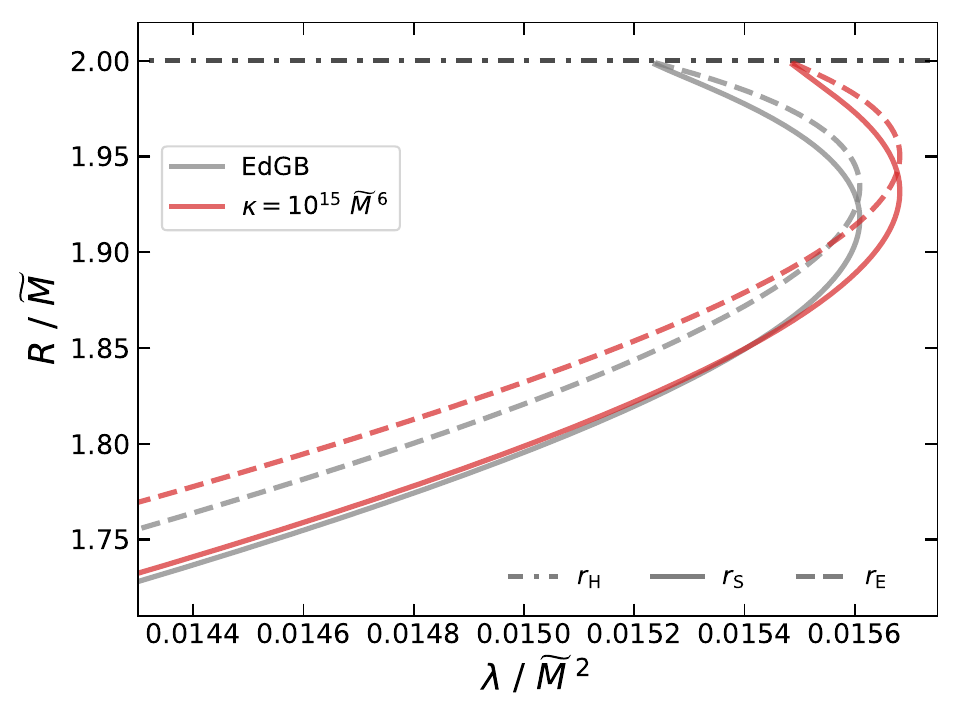}
    \caption{Position of the curvature singularity and breakdown of hyperbolicity for spherically symmetric BH solutions in $f(\RR)$-dGB gravity. We show the quadratic case in the upper panel, with different colors referring to different choices of the constant $\kappa$, and the quartic case in the bottom panel. In both plots the gray lines refer to the EdGB case. The dot-dashed line denotes the position of the horizon which is the same for all the solutions we constructed; solid lines denote the position of the curvature singularity, where the Kretschmann scalar diverges; dashed lines denote the radius $\rE$ where strong hyperbolicity breaks down. Interestingly, despite the structure of the system of equations being radically different from the EdGB case, the qualitative picture looks significantly similar.}
    \label{fig:HyperbolictyBreakdown}
\end{figure}

Our results for the breakdown of strong hyperbolicity on static and spherically symmetric BH solutions are shown in Fig.~\ref{fig:HyperbolictyBreakdown}, where the top panel refers to the quadratic case ($n = 2$), while the bottom panel to the quartic one ($n = 4$). In both plots the solid lines denote the position of the singularity while the dashed ones denote the radius $\rE$. The gray lines denote the EdGB case and, as we can see, for both choices of $f(\RR)$ considered, the qualitative picture is strikingly similar. This is remarkable as, even though the mass-radius diagram shows small deviations from the EdGB case, the structure of the field equations is significantly altered, with new degrees of freedom appearing and the determinant of the principal symbol becoming a fourth degree polynomial. Furthermore, as previously discussed, the BH interior region is significantly affected by the $f(\RR)$ term, although this does not remove the elliptic region.

\section{Conclusions}
\label{sec:Conclusions}

We have explored an extension of GR that combines $f(\RR)$ gravity and EdGB gravity into a unified framework featuring two nonminimally coupled scalar degrees of freedom. This construction goes beyond Horndeski's class and allows for the inclusion of arbitrary higher-curvature terms at the nonperturbative level, preserving second-order field equations and providing a versatile framework to investigate the UV behavior of gravity and the fate of singularities.

Focusing on quadratic and quartic curvature corrections, we have shown that the interplay between $f(\RR)$ and Gauss–Bonnet terms leads to genuinely new phenomena that are absent in either theory taken separately. In particular, $f(\RR)$ corrections alter BH geometries, while the qualitative features characteristic of EdGB BHs ---such as the existence of a minimum mass and multiple solution branches --- are preserved. We have also identified a suppression mechanism for the divergence of the Ricci scalar inside the horizon, while the overall singularity structure and the emergence of elliptic regions remain akin to those of pure EdGB gravity, even in the presence of higher-curvature $f(\RR)$ terms. This suggests that adding isolated higher-order curvature operators does not, by itself, cure the well-posedness issues inherent to EdGB-type theories at the nonperturbative level.

Different strategies to cure potential pathologies arising in UV extensions of GR include adding higher-order operators within the context of effective field theories~\cite{Figueras:2024bba,Figueras:2025gal}, or an \emph{infinite} tower of curvature invariants~\cite{Bueno:2024dgm,Bueno:2025zaj}.

Our findings highlight both the potential and the limitations of higher-curvature extensions of gravity when considered at the full nonperturbative level. Future work should address the nonlinear dynamics of this theory, including the formation and evaporation of BHs. However, based on our results, we expect to find a qualitatively similar phenomenology as in EdGB in dynamical settings: elliptic regions will form during the gravitational collapse~\cite{Ripley:2019hxt,Ripley:2019irj}, in BH mergers~\cite{East:2020hgw,East:2021bqk,Corman:2022xqg}, or when simulating Hawking radiation past the minimum mass~\cite{Corelli:2022phw,Corelli:2022pio}.

Concerning the feasibility of numerical simulations in $f(\RR)$-dGB theory, some technical aspects are worth considering.
The shooting procedure we adopted to construct static BH solutions requires an extremely high accuracy to prevent numerical divergences at the boundaries. In particular, for the quartic case 55 significant figures were required to set the outer boundary of our computation domain to $r_\infty = 80 \div 130 \, \M$. In a numerical simulation one might wish to push $r_\infty$ to larger values, which would require an even higher accuracy, and a higher computational cost. Nevertheless, the same precision might not be required for time evolution, and using the common floating point data types of languages like C/\cpp{} might not result in numerical instabilities. In this regard, it would be interesting to check if even the initialization procedure can be made more efficient by employing a relaxation method (see, e.g.,~\cite{Sennett:2017etc} for the case of boson stars with large self-interactions).

\section*{Acknowledgments}
This work is supported by the MUR FIS2 Advanced Grant ET-NOW (CUP:~B53C25001080001) and by the INFN TEONGRAV initiative. A. P. S. also gratefully acknowledges support from a research grant funded under the INFN–ASPAL agreement as part of the Einstein Telescope training program.

\bibliographystyle{apsrev4-1}
\bibliography{Refs.bib}

\end{document}